\documentclass[prl,twocolumn,preprintnumbers,superscriptaddress,showpacs]{revtex4-1}
\usepackage{graphicx}
\usepackage{amsmath}
\usepackage{amssymb}
\usepackage{xcolor}

\newcommand\sect[1]{\vspace{.15cm} \noindent \textbf{#1} -- }
\newcommand*{\Ang}{\ensuremath{{\mbox{\normalfont\AA}}}}
\newcommand*{\eV}{\ensuremath{{\mathrm{eV}}}}

\renewcommand{\u}{\ensuremath{{\mathbf{u}}}}
\newcommand{\w}{\ensuremath{{\mathbf{w}}}}
\renewcommand{\P}{\ensuremath{{\mathbf{P}}}}
\newcommand{\B}{\ensuremath{{\mathbf{B}}}}
\renewcommand{\k}{\ensuremath{{\mathbf{k}}}}

\usepackage{pdfpages}
\makeatletter
\AtBeginDocument{\let\LS@rot\@undefined}
\makeatother

\makeatletter

\newcommand*\wthelper[2]{%
    \hbox{\dimen@\accentfontxheight#1%
        \accentfontxheight#11.15\dimen@
        $\m@th#1\widetilde{#2}$%
        \accentfontxheight#1\dimen@
    }%
}
\newcommand*\accentfontxheight[1]{%
    \fontdimen5\ifx#1\displaystyle
        \textfont
    \else\ifx#1\textstyle
        \textfont
    \else\ifx#1\scriptstyle
        \scriptfont
    \else
        \scriptscriptfont
    \fi\fi\fi3
}
\makeatother

\begin{document}

\author{Jing-Yuan~Chen}
\affiliation{Stanford Institute for Theoretical Physics, Stanford University, Stanford, CA 94305, USA}
\author{Steven~A.~Kivelson}
\affiliation{Department of Physics, Stanford University, Stanford, CA 94305, USA}
\author{Xiao-Qi~Sun}
\affiliation{Department of Physics, Stanford University, Stanford, CA 94305, USA}
\affiliation{Stanford Center for Topological Quantum Physics, Stanford University, Stanford, CA 94305, USA}

\title{Enhanced thermal Hall effect in nearly ferroelectric insulators}
\date{\today}

\begin{abstract}
In the context of recent experimental observations of an unexpectedly large thermal Hall conductivity, $\kappa_H$, in insulating $\mathrm{La_2CuO_4}$ (LCO) and $\mathrm{SrTiO_3}$ (STO), we theoretically explore conditions under which acoustic phonons can give rise to such a large $\kappa_H$. Both the intrinsic and extrinsic contributions to $\kappa_H$ are large in proportion to the dielectric constant, $\epsilon$, and the ``flexoelectric'' coupling, $F$. While the intrinsic contribution is still orders of magnitude smaller than the observed effect, an extrinsic contribution proportional to the phonon mean-free path appears likely to account for the observations, at least in STO. We predict a larger intrinsic $\kappa_H$ in certain insulating perovskites.
\end{abstract}

\maketitle

\sect{Introduction} While it is well known that ``neutral'' excitations in solids, including phonons and other collective modes, induce some charge motion, it is intuitively clear that this is in some sense a ``small'' effect. In particular, this suggests that the coupling of such modes to magnetic fields is generically weak, and consequently that their contribution to the Hall component of the thermal conductivity tensor, $\kappa_H$, is relatively small. This argument is assumed implicitly when the ratio of $\kappa_H$ to the Hall conductivity is used to test the Weidemann-Franz law in metals. It is also why the recent observation of a 
large $\kappa_H$ 
in $\mathrm{La_2CuO_4}$ (LCO) \cite{grissonnanche2019giant}, an insulating cuprate, 
generated so much interest \cite{samajdar2019thermal,samajdar2019enhanced,han2019consideration,li2019thermal}. 
Moreover, an anomalous 
contribution to $\kappa_H$ of 
smaller but still comparable magnitude has been identified in the doped material, $\mathrm{La_{2-{\it x}}Sr_{\it x}CuO_4}$ (LSCO), for a range of $0\leq x < 0.19$ comprising much of the ``high temperature'' superconducting range of doping. Still more recently, a 
comparably large $\kappa_H$ has been found in the nearly ferroelectric insulator $\mathrm{SrTiO_3}$ (STO) \cite{li2020phonon}.

In this letter we analyze the contribution of phonons to $\kappa_H$ at temperatures small compared to the Debye temperature, so as to identify conditions under which it can be larger than expected on the basis of dimensional analysis. Naively, $\kappa_H$ is small compared to the longitudinal response, $\kappa_L$, for two reasons: {\bf i) } $\kappa_H$ is small in proportion to $B/B_0$ where $B$ is the applied field, and $B_0 \equiv \phi_0/a^2 \sim 10^4\mathrm{T}$ is the magnetic field corresponding to one quantum of flux ($\phi_0= 2\pi\hbar c/e$) per unit cell crossectional area, $a^2$. {\bf ii)} $\kappa_L$ is large in proportion to $\ell/a$, the ratio of the phonon mean free path to the lattice constant. However, especially in the context of STO, we show that $\kappa_H$ is enhanced by a factor proportional to the dielectric constant, $\epsilon$, times the flexoelectric coupling $F$ (defined below). 
Moreover, we find that 
there is an extrinsic contribution to $\kappa_H$ that -- in common with $\kappa_L$ -- is proportional to $\ell$. A combination of these effects is the likely explanation of the large $\kappa_H$ observed in STO; we speculate that they are responsible for the anomalous thermal Hall response in the cuprates, as well.
 
To be explicit, at low temperatures in an insulator, the dominant heat carriers are the acoustic phonons. 
We can express the thermal conductivity tensor in terms of the thermal diffusivity $D$ as
\begin{align}
\kappa^{ij} = C_v\: D^{ij}
\label{thermal_diffusivity}
\end{align}
where $i, j=x, y$, $C_v =(2\pi^2/5) (T/\hbar v_1)^3$ is the specific heat per unit volume, $v_1$ is an appropriate average of the sound speed over polarizations and directions of propagation, and we will use units in which $k_B=1$. As is well known, the longitudinal piece of ${\bf D}$ is $D_L\equiv D^{xx} = D^{yy} = (1/3) v_2 \ell$ where $v_2$ is a slightly different average of the sound speed \cite{beck1974phonon}. In the following, we will focus on $D_H \equiv (1/2)(D^{xy}-D^{yx})$. 
\footnote{Here $\kappa_H$ is defined 
as the thermal Hall conductance per unit thickness in the $\hat{\mathbf{z}}$ direction. 
 For $\kappa_H$, 
 in contrast with the electronic $\sigma_H$, the macroscopically averaged value and the bulk local value are drastically different, as there is a substantial cancellation between the heat current carried in the bulk and near the edge \cite{cooper1997thermoelectric, qin2011energy,  qin2012berry}.} 
%
Here we have assumed $\B = B\hat{\mathbf{z}}$ and assumed rotational symmetry at low energies for simplicity; our results will be qualitatively unchanged in lower symmetry situations. 

%

\sect{Effective field theory} The low energy dynamics in a nearly ferroelectric insulator is described by the continuum Lagrangian density $L=L_0 + L_{int} + L_{P}$ in terms of the vector fields $\u$ and $\P$, which represent the local acoustic displacement and the dipole density respectively.
 
$L_0$ is the bare Lagrangian of the acoustic modes
 \begin{align}
L_0=\frac{\rho}{2} \dot \u^2-\frac{K_1}{2}(\nabla \u)^2-\frac{K_2}{2}(\nabla\cdot \u)^2 +\ldots
\label{L0}
\end{align}
where $\rho$ is the mass density, and $K_a$ the elastic moduli.
Here and below ``$\ldots$'' refers to higher derivative terms.

Near ferroelectric quantum criticality \cite{rowley2014ferroelectric}, while there is no net dipole density $ \langle \P\rangle$, it is essential to include a fluctuating dipole order parameter $\P$ which arises from a combination of all the infra-red active phonon modes.
To leading order 
\begin{align}
L_P = -\frac{\P^2}{2\chi\epsilon_0} + \ldots
\end{align}
where $\chi\equiv \epsilon/\epsilon_0-1$ is the static electric susceptibility. Important subleading dynamical terms are discussed later with Eq.\eqref{L_w}.

Finally, the terms that couple the acoustic displacement to the dipole density take the form
\begin{align}
L_{int} = \: & \frac{\B}{c} \cdot \left(\P\times \dot{\u}\right) \nonumber \\[.1cm]
& + F_1 \: \P\cdot \nabla^2 \u + F_2 \: \P\cdot \nabla(\nabla\cdot \u) + \ldots \ .
\end{align}
These terms 
are the lowest order terms (in powers of fields and derivatives) allowed by symmetry. 
The couplings $F_a$ are known as flexoelectric couplings \cite{zubko2013flexoelectric}. Their meaning is 
clear from the corresponding ``adiabatic'' equations of motion for $\P$, valid so long as the acoustic displacements are slowly varying 
compared to the energy scale of the lowest optical modes (see Eq.\eqref{L_w})
\begin{align}
\frac{\P}{\chi \epsilon_0} = \dot \u \times \frac{\B}{c} + F_1 \nabla^2 \u + F_2 \nabla(\nabla \cdot \u) + \ldots \ .
\label{P_EoM}
\end{align}
The first term is the attractive force between the two poles of a moving dipole in magnetic field. The two flexoelectric terms represent the dipole density induced by a strain gradient.

\sect{Intrinsic thermal Hall conductivity} While in the absence of scattering, $\kappa_{L}$ diverges, 
there is a well-defined intrinsic thermal Hall conductivity \cite{qin2012berry}.
Here, we integrate out $\P$ using Eq.\eqref{P_EoM} for an effective Lagrangian $L_\mathrm{eff} = L_0 + L_B$, which (to linear order in $B$) modifies the dynamics of the acoustic mode by
\begin{align}
L_B =\chi\epsilon_0\ \frac{\B}{c} \cdot 
 \left[
F_1 \nabla^2 \u + F_2 \nabla(\nabla \cdot \u) \right] \times \dot\u
+ \ldots \ .
\label{L_B}
\end{align}
Although the induced $L_B$ is ``small'', both due to the $B$ dependence and the higher derivatives on $\u$, it is nevertheless the leading time-reversal odd term and therefore important.
In particular, it introduces a time-reversal odd contribution to the Berry curvature $\mathbf\Omega_\alpha(\k)$, where $\alpha$ labels the phonon mode and $\k$ its momentum. The resulting anomalous motion of the phonons generates an intrinsic thermal Hall conductivity \cite{qin2012berry}:
\begin{align}
\kappa_H^{in} = T\sum_{\alpha} \int\frac{d^3 k}{(2\pi\hbar)^3} \: \Omega^z_\alpha(\k) \int_{{E_\alpha(\k)}/{T}}^\infty d\xi \: \xi^2 \frac{\partial f_{BE}(\xi)}{\partial \xi}
\end{align}
where $f_{BE}(\xi)$ is the Bose-Einstein distribution at energy $E=\xi T$. 

A phonon effective Lagrangian of the present form has been studied in Ref.~\cite{qin2012berry}. The computation of the Berry curvature is reproduced in the Supplemental Material. Its time-reversal odd components are found to be of order
\begin{align}
\Omega(\k) \sim \frac{\chi\epsilon_0 F B vk}{Kc \hbar} \frac{\hbar}{k^2}
\label{Omega}
\end{align}
($k\equiv |\k|$, and $v \equiv \sqrt{K/\rho}$ is the sound velocity) 
where $K$ and $F$ are characteristic values of the couplings $K_1$, $K_2$ and $F_1$, $ F_2$ respectively.  
The parameter dependence of $\Omega$ is physically intuitive: The first, dimensionless, factor is the typical ratio between $L_B$ and $L_0$ (in the spirit of perturbation theory in $B$), where we have estimated $\partial/\partial t$ as $vk/\hbar$; the second factor is because $\Omega$ is defined with two $k$-derivatives times $\hbar$. The same kind of reasoning 
yields the estimate
\begin{align}
\frac{\kappa_H^{in}}{\kappa_L} \sim \frac{\chi\epsilon_0 F B T}{Kc \hbar} \frac{v\hbar}{T \ell} \ ,
\label{kappain_ratio}
\end{align}
where the second dimensionless factor is because $\kappa_L$, given below Eq.\eqref{thermal_diffusivity}, scales with $\ell$. This analysis also implies $\kappa_H^{in} \propto T^3$ as $\kappa_L$ does \cite{qin2012berry}. 

In a nearly ferrolelectric material, there is a softened optical phonon branch $\w$ that contributes dominantly to $\P$ \cite{rowley2014ferroelectric}. 
{ At} temperatures comparable to or above the optical phonon gap,  
the dipole density $\P = \rho_e \w + \ldots$ , where $\rho_e$ is roughly the charge on the positive ions per unit cell, and ``$\ldots$'' includes contributions from the flexoelectric effect Eq.\eqref{P_EoM} as well as possibly higher optical phonons. In this case, one must include the heat carried by $\w$. The effective Lagrangian should include the extra terms
\begin{align}
& - \frac{\rho' }{2\hbar^2}\Delta_{\text{TO}}^2 \w^2 + \frac{\rho'}{2} \dot\w^2 - \frac{K'_1}{2}(\nabla \w)^2-\frac{K'_2}{2}(\nabla\cdot \w)^2 +\ldots \nonumber \\[.1cm]
& + \frac{\B}{c} \cdot \left(\rho_e \w \times \dot{\u}\right) + \frac{\B}{c} \cdot \left(\rho'_e \w \times \dot{\w}\right) + \ldots
\label{L_w}
\end{align}
where $\Delta_{\text{TO}}$ is the transverse optical phonon energy and 
{ the ``$\ldots$'' includes the long-range Coulomb interaction \cite{lyddane1941,ruhman2016} that causes an upward shift of the longitudinal optical phonon energy. }
Compared to the acoustic phonons, while the thermally excited $\w$ phonons are fewer in number due to the gap, their couplings to $\B$ 
{ their couplings are larger in  the sense  that they} involve fewer derivatives than in $L_B$. Therefore, at finite temperatures, 
corrections to $\kappa_H^{in}$ due to $\w$ should be considered. The details of both the $\u$ and the $\w$ contributions to $\kappa_H^{in}$ are presented in the Supplemental Material.

While the intrinsic thermal Hall effect has been concretely studied, it is most often negligibly small in real materials. As we will see later, in STO, even after an enhancement by the low temperature electric susceptibility $\chi \approx 2\times 10^4$ \cite{dec1999scaling}, $\kappa_H^{in}$ is still $10^{-4}$ smaller than the observed value. We will however predict candidate materials in which this intrinsic effect might be sufficiently large for observation.

\sect{Extrinsic thermal Hall conductivity} Skew scattering plays an important role in the Hall conductivity $\sigma_H$, most notably for its linear in $\ell$ 
contribution \cite{smit1955spontaneous, smit1958spontaneous, majumdar1973hall, nagaosa2010anomalous}. Now we show there is a parallel effect in the phonon thermal Hall conductivity $\kappa_H$, under appropriate, but unparallel, 
conditions.

There are two origins of time-reversal oddness during a scattering event. 
The first is directly associated with the defect off which the particle scatters, and second the Berry curvature effects on the kinetics of the particle itself. The former is the mechanism responsible for the $\ell$ linear contribution to the electronic $\sigma_H$ \cite{nagaosa2010anomalous}. Here for phonon $\kappa_H$, we focus on the latter, i.e. Berry curvature induced skew scattering, primarily because in STO there is no evidence of magnetic defects.

We assume 
dilute defects that scatter 
phonons strongly. To understand the 
importance of this assumption, 
first recall the opposite case of 
weak scattering. In this case, the disorder averaged $\overline{(\Delta H)^2}$ is a small parameter, where $\Delta H$ is the modification of Hamiltonian density by the disorder. As a consequence of Fermi's golden rule, the typical scattered angle at each event, $\Delta \hat\k$, and the inverse of the mean free path due to accumulated mild events, $1/\ell$, are both of smallness $\overline{(\Delta H)^2}$. Since Fermi's golden rule is manifestly time-reversal even, skew scattering only happens at higher orders in $\Delta H$, which would be $\overline{(\Delta H)^4} \sim 1/\ell^2$ if the time-reversal oddness is solely due to the particle but not the defect. This is smaller than the non-skew scatterings by $1/\ell$; as a consequence, in electronic systems \cite{nagaosa2010anomalous}, the contribution of Berry curvature induced skew scattering to $\sigma_H$ does not scale with $\ell$. 
\footnote{If $\overline{\Delta H} \neq 0$ in electron scattering (e.g. when disorders are mostly repulsive \cite{ishizuka2017noncommutative}), then the leading Berry curvature induced skew scattering occurs at $\overline{(\Delta H)^3} \sim 1/\ell^{3/2}$, hence its contribution to $\sigma_H$ is $1/\ell^{1/2}$ smaller than $\sigma_L$.}
On the other hand, for strong scattering, 
Fermi's golden rule does not apply. The typical scattered angle $\Delta\hat\k$ at an event is of order $\pi$, and the mean free path $\ell$ is about the actual spacing between the defects; neither scales with the indefinitely large $\Delta H$. Thus, the Berry curvature induced skew scattering 
 scales as $1/\ell$ along with non-skew scattering. For phonons specifically, a smoking-gun for strong scattering is 
 that $\ell$ approaches a finite constant as $T\to 0$. 
(This is ``boundary-like'' scattering \cite{casimir1938note, guyer1966solution, guyer1966thermal, beck1974phonon}, 
which we take to be strong and approximately independent of the phonon energy at $B=0$.) This will be associated with STO phenomenology later.

With this physical picture in mind, we can write down a linearized Boltzmann equation describing phonon transport in the presence of a static temperature gradient, 
 including an ansatz for the Berry curvature induced skew scattering with dimensionless strength $A$:
\begin{align}
 v \hat\k \cdot \nabla\delta T \frac{\partial f_{BE}(\xi)}{\partial T} = \ -\frac{\delta f(\k) + \delta' f(\k)}{\tau} \hspace{1cm} \nonumber \\[.1cm]
 +\int_{k'=k} d^2\hat{k'} \ \frac{{A}}{\hbar\tau} \ \mathbf{\Omega}(k) \cdot (\k \times \k') \nonumber \\[.1cm]
\ \cdot \ \delta f(\k') [1+2f_{BE}(\xi)]
\label{Boltz_Eq}
\end{align}
 Here we separated the distribution $f=f_{BE} + \delta f + \delta' f$, where $\delta f$ is of smallness $\nabla\delta T$, and $\delta' f$ of smallness $B\, \nabla\delta T$, and kept $\nabla\delta T$ and $B$ each to linear order; $\tau=\ell/v$ is the relaxation time, and $\xi\equiv vk/T$. In writing the skew scattering ansatz, for simplicity we have assumed it is dominated by elastic single phonon scattering, whose full collision kernel is
\begin{align}
\int d^3 k' \ \delta(vk-vk') \ \left\{\ W_{\k'\rightarrow \k}\ f(\k') [1+f(\k)] \right. \ \ \ \ \nonumber\\[.1cm]
\left. - W_{\k\rightarrow\k'}\ f(\k) [1+f(\k')]\ \right\}. 
\end{align}
Time reversal transformation requires $W_{\k' \rightarrow \k}(\B) = W_{-\k\rightarrow -\k'}(-\B)$ in the scattering probability. The zeroth order in $B$ non-skew scattering contributes to the relaxation time approximation (among other multi-phonon processes), while the linear in $B$, Berry curvature induced, skew scattering is approximated by our ansatz in Eq.\eqref{Boltz_Eq}. Importantly, the dimensionless parameter $A$ approaches a constant at small $k$, as we will justify later.

The Boltzmann equation Eq.\eqref{Boltz_Eq} is easily solved by matching orders in $B$:
\begin{align}
\delta f(\k) &= \ell \: \hat\k\cdot\nabla\delta T \ \frac{\xi}{T} \frac{\partial f_{BE}(\xi)}{\partial \xi}, \nonumber \\[.1cm]
\delta' f(\k) &= \ell {A} \: \frac{\chi\epsilon_0 F v}{K\hbar} \frac{\B}{c} \cdot \frac{\k \times \nabla\delta T}{3} \nonumber \\[.1cm]
& \hspace{1.5cm} \cdot \frac{\xi}{T} \frac{\partial f_{BE}(\xi)}{\partial \xi} [1+2f_{BE}(\xi)] \ ,
\end{align}
where we have used Eq.\eqref{Omega} for $\mathbf\Omega$, with possible order $1$ factors absorbed into ${A}$. We can then substitute these solutions into the heat current $\mathbf{J} = \int[d^3 k /(2\pi\hbar)^3] \: v\hat\k \: vk \left[ \delta f(\k) + \delta' f(\k)\right]$ and match with $\mathbf{J} = \kappa_L (-\nabla \delta T) + \kappa_H (-\nabla \delta T) \times \hat{\mathbf{z}}$ to identify $\kappa_L$ and $\kappa_H$, which are associated with $\delta f$ and $\delta' f$ respectively. At this point, we shall restore the fact that there are three acoustic modes, hence a multiplicity of $3$ for $\kappa_L$ and a multiplicity of $9$ for $\kappa_H$ (the extra $3$ comes from $\sum_{\alpha'} \delta f_{\alpha'}(\k')$ in the collision). This leads to $\kappa_L$ given below Eq.\eqref{thermal_diffusivity}, and
\begin{align}
\kappa_H^{ex} \sim & \: \ell {A} \left(\frac{T}{\hbar v}\right)^3 T \: \frac{\chi\epsilon_0 F B v}{Kc \hbar} \nonumber \\[.2cm]
& \cdot \frac{4\pi}{(2\pi)^3} \int_0^\infty d\xi \: \xi^5 \frac{-\partial f_{BE}(\xi)}{\partial \xi} [1+2f_{BE}(\xi)].
\end{align}
The integral can be readily performed, and the result is most conveniently presented as the ratio
\begin{align}
\frac{\kappa_H^{ex}}{\kappa_L} \sim 5{A} \: \frac{\chi\epsilon_0 F B T}{Kc \hbar}.
\label{kappaex_ratio}
\end{align}
This is our 
{ principal result for} the extrinsic thermal Hall effect. Since $\kappa_L\propto T^3$, we predict $\kappa_H^{ex}\propto T^4$. Compared to the intrinsic effect Eq.\eqref{kappain_ratio}, we have an 
enhancement by $5{A} T\ell /v\hbar$. We will see this linear in $\ell$ enhancement (with a phenomenological parameter ${A}$ of order $1$), together with the large $\chi$, reproduces the observed large $\kappa_H$ in STO.

It remains to justify the claim that ${A}$ approaches a constant at low energies, which is crucial for the $T^4$ scaling of $\kappa_H^{ex}$. Note that the dimensionless factor ${A}\mathbf\Omega\cdot(\k\times \k')/\hbar$ in our ansatz Eq.\eqref{Boltz_Eq} is the time reversal odd modification to the scattering probability from $\hat\k'$ to $\hat\k$. It is controlled by the typical ratio between $L_B$ and $L_0$, with the extra $\partial/\partial t$ in $L_B$ estimated as $vk/\hbar$. Importantly, this $\partial/\partial t$ should not be estimated as $\Delta H/\hbar$ even though we are considering it during the course of a scattering event, because as we explained before, in strong scattering events no scattering angle would scale with the indefinitely large $\Delta H$. Hence ${A}\mathbf\Omega\cdot(\k\times \k')/\hbar$ scales as $k$, and therefore, combined with Eq.\eqref{Omega}, ${A}$ approaches a constant, whose value is determined by but does not scale with $\Delta H$.

Finally, 
there is another extrinsic contribution to Hall physics, that a particle's center of wavepacket experiences a ``side-jump'' by $\Delta \mathbf{r} \sim \mathbf\Omega \times \Delta \k$ during the course of scattering \cite{nagaosa2010anomalous}. This extrinsic effect, unlike the skew scattering, does not scale with $\ell$ and hence is of the same order as $\kappa_H^{in}$. The reason is intuitive: while the shifted distance due to a modified (skew) scattering angle increases with propagation time, the shift that happened at the instant of scattering does not. 

\sect{Discussion} The behavior of $\kappa_H$ in STO reported in Ref.~\cite{li2020phonon} has several 
important features, in addition to the surprisingly large magnitude. Firstly, $\kappa_H$ is peaked at approximately the same temperature, $T_{peak} \approx 20\mathrm{K}$, as $\kappa_L$ 
 -- which certainly suggests \cite{li2020phonon} that they 
 both reflect heat transport by the same set of excitations; 
in particular, above 
$T_{peak}$, acoustic phonons start to 
lose momentum by Umklapp scattering \cite{beck1974phonon, guyer1966solution, guyer1966thermal}. 
Moreover, the recently measured $\kappa_L$ scales as $T^3$ at low temperatures $\lesssim 10\mathrm{K}$ \cite{martelli2018thermal}, and $\kappa_H$ scales as $T^4 B$ in the same temperature range \cite{li2020phonon}. The $T^3$ scaling of $\kappa_L$ implies the phonon mean free path $\ell$ is roughly $T$ independent at low temperatures, and is extracted according to Eq.~\eqref{thermal_diffusivity} to be $\sim 1\mu\mathrm{m}$. 
\footnote{Between $10-20\mathrm{K}$ the phenomenon of Poiseuille flow \cite{guyer1966solution, guyer1966thermal} is observed \cite{martelli2018thermal}, where $\kappa_L$ increases faster than $T^3$ with $T$. This is because the onset of phonon-phonon interactions decreases the true diffusion velocity compared to the ballistic phonon velocity, and hence increases the relaxation time. Similar increase is observed in $\kappa_H$ in the same temperature range \cite{li2020phonon}, in support of our interpretation of $\kappa_H$ as an extrinsic effect.}
%
The temperature independent $\ell$ 
has been interpreted \cite{martelli2018thermal, li2020phonon} as the scattering of phonons off the twin boundaries between tetragonal domains in STO \cite{buckley1999twin}. 
Importantly, the scattering must not be soft refraction and reflection, but rather some strong interaction process with the localized degrees of freedom on the twin boundaries (such as the localized dipoles \cite{scott2012domain}), 
in order to produce a temperature independent $\ell$ \cite{casimir1938note, guyer1966solution, guyer1966thermal, beck1974phonon} comparable to the twin boundary spacing. While a detailed scattering mechanism has yet to be established, these known qualitative aspects 
all support the applicability of our theory of extrinsic thermal Hall effect. 

To begin with, the observed $\kappa_H \sim T^4$ scaling is consistent with that predicted in Eq.\eqref{kappaex_ratio}. Our theory 
reproduces the observed magnitude of $\kappa_H$ with the dimensionless parameter ${A}$ of order $1$. STO 
has a large $\chi \approx 2\times 10^4$ at low temperatures \cite{dec1999scaling}. On the other hand, the flexocouplings take ``normal'' values $F \approx e/4\pi\epsilon_0 a$ \cite{kogan1964piezoelectric} (where $a=3.9\Ang$), according to experiments \cite{zubko2007strain, zubko2008erratum} and numerics \cite{hong2010flexoelectricity}. The elastic moduli components take the usual values $K\approx1\eV/\Ang^3$ \cite{scott1997interpretation}.
Thus, at $B/c=10\mathrm{T}=1.5\times 10^{-4} \hbar/e\Ang^2$ and $T=10\mathrm{K}= 8.6\times 10^{-4}\eV$, our theory Eq.\eqref{kappaex_ratio} 
yields $\kappa_H^{ex}/\kappa_L \approx 5{A} \times 5\times 10^{-5}$. This 
matches the observed $\kappa_H/\kappa_L \sim -10^{-3}$ at this magnetic field and temperature, 
 %
 \footnote{The measured values of $\kappa_H$ appear to vary by of order a factor of two for different samples, and even for the same sample if a thermal cycle is performed across the tetragonal transition temperature at $105\mathrm{K}$. This is taken as evidence that $\kappa_H$ is due extrinsically to twin boundaries \cite{li2020phonon}, although why the dependence on thermal history is larger in $\kappa_H$ than in $\kappa_L$ remains unclear without a modeling of the scattering process.}
if we set ${A} \approx 4$. (The sign is inconclusive because the component-wise values and signs of $F$ remain unsettled \cite{zubko2013flexoelectric}.)
\footnote{We have used the bulk value of $\chi \epsilon_0 F$. Near the twin boundaries where the scatterings occur, the local value is likely different (probably enhanced) due to the localized dipoles and structures \cite{zubko2007strain, scott2012domain, morozovska2012impact}.}
%
On the other hand, the intrinsic thermal Hall effect Eq.\eqref{kappain_ratio} gives a ratio $\kappa_H^{in}/\kappa_L \approx 2\times 10^{-7}$ for $\ell\sim 1\mu\mathrm{m}$, much smaller than the observed value. We estimate that the contribution from the soft optical phonon $\w$ (with $\Delta_{\text{TO}} \approx 24\mathrm{K}$ \cite{rowley2014ferroelectric}) { at $T= 10$K is 
less than} $0.1$ of that from acoustic phonon, see Supplemental Material.

According to our theory, 
a change in the mean free path is not expected to dramatically change $\kappa_H/\kappa_L$. 
However, if a single crystal domain is formed (e.g. by cooling under strain \cite{li2020phonon}), although $\l$ becomes the system size, skew scattering ceases to happen during the transport, and hence $\kappa_H$ should drop to the order of $\kappa_H^{in}$. This is consistent with the negligible $\kappa_H$ in $\mathrm{KTaO_3}$ (KTO) \cite{li2020phonon}. Note that $\ell$ in KTO is determined by the system size below $1\mathrm{K}$ and by soft impurity scattering at higher temperatures \cite{salce1994disorder}, and indeed neither case 
{ produces} an $\ell$ linear $\kappa_H^{ex}$ according to our theory.

While the intrinsic $\kappa_H^{in}$ is negligible in STO even with the large $\chi \approx 2\times 10^4$, we predict $\kappa_H^{in}$ to be observable in other systems, in particular a large class of perovskites \cite{zubko2013flexoelectric}, including $\mathrm{Ba_{1-{\it x}}Sr_{\it x}TiO_3}$ (BSTO) and $\text{PMN-PT}$, which not only have similarly large $\chi$ but also large flexocouplings $F$ hundreds of times of the ``normal'' value $e/4\pi\epsilon_0 a$. Furthermore, if some of these 
materials form structural domains at low temperatures, their $\kappa_H^{ex}$ should be significantly larger than in STO.

Finally, we turn to the observations in LCO and LSCO. 
In Ref.~\cite{grissonnanche2019giant}, after 
subtracting off a quasiparticle contribution (inferred from $\sigma_H$ using the Weidemann-Franz law), 
the remaining ``anomalous'' contribution is found to decrease smoothly with hole doping $x$, extending to values of $x$ greater than the ``optimal doping'' that maximizes $T_c$; indeed, it is suggested it vanishes only for $x > p^\star$,
an independently determined (material dependent) crossover concentration that is roughly $p^\star \approx 0.19$ in LSCO. 
In the range $0\leq x <p^\star$, the system evolves from an antiferromagnetically ordered insulator 
through an insulating spin-glass phase 
and over much of the superconducting dome.
The only low energy excitations that exist over this entire range of doping are the acoustic phonons. 
 
While in LCO, which is far from ferroelectricity, the electric susceptibility $\chi \approx 30$ \cite{chen1991frequency} is much smaller than in nearly ferroelectric materials such as STO, the flexocouplings $F$ are unknown and could potentially be larger 
as in 
BSTO. Moreover, in LSCO, there is a $T=0$ structural transition (from orthorhombic to tetragonal) at $x \approx p^\star$, which might be significant in the context of a phonon skew scattering mechanism. 
The skew scattering might also originate from 
magnetic defects, the existence of which is plausible given that LCO is an antiferromagnet, and that spin-glass order persists up to $x \approx p^\star$ \cite{frachet2019hidden} (albeit at lower $T$ and larger $B$ than those in Ref.~\cite{grissonnanche2019giant}). 
On the other hand, in LCO at the lowest $T$ probed so far, $\kappa_L $ (as well as $\kappa_H$) drops roughly linearly in $T$; 
this is not the behavior expected in the transport regime we have explored, proving that the present analysis is not directly applicable. 
Nonetheless, the fact that a phonon mechanism can produce a thermal Hall response of the requisite size in STO leads us to conjecture that the same physical considerations are
at play in 
LCO as well.

\sect{Note added} While this paper was under review, a new paper \cite{newLouis} appeared, reporting a comparably large thermal Hall effect in LCO when the thermal current is oriented perpendicular to the Cu-O planes.  As noted by the authors, this observation of isotropy establishes beyond reasonable doubt that the thermal Hall current in LCO is carried by acoustic phonons.
\vspace{.15cm}

\acknowledgements
We thank Kamran~Behnia and Xiaokang~Li for communications of their experimental results, Tao~Qin and Junren~Shi for discussions on phonon Berry curvature, John~Tranquada, Louis~Taillefer, and Jonathan~Ruhman for helpful comments. J.-Y.~C. is supported by the Gordon and Betty Moore Foundation's EPiQS Initiative through Grant GBMF4302. S.~A.~K. is supported in part by the U. S. Department of Energy (DOE) Office of Basic Energy Science, Division of Materials Science and Engineering at Stanford under contract No. DE-AC02-76SF00515. X.-Q.~S. is supported by the DOE Office of Science, Office of High Energy Physics, under the contract No. DE-SC0019380.

\bibliography{phonon_ThHall}

\begin{thebibliography}{41}%
\makeatletter
\providecommand \@ifxundefined [1]{%
 \@ifx{#1\undefined}
}%
\providecommand \@ifnum [1]{%
 \ifnum #1\expandafter \@firstoftwo
 \else \expandafter \@secondoftwo
 \fi
}%
\providecommand \@ifx [1]{%
 \ifx #1\expandafter \@firstoftwo
 \else \expandafter \@secondoftwo
 \fi
}%
\providecommand \natexlab [1]{#1}%
\providecommand \enquote  [1]{``#1''}%
\providecommand \bibnamefont  [1]{#1}%
\providecommand \bibfnamefont [1]{#1}%
\providecommand \citenamefont [1]{#1}%
\providecommand \href@noop [0]{\@secondoftwo}%
\providecommand \href [0]{\begingroup \@sanitize@url \@href}%
\providecommand \@href[1]{\@@startlink{#1}\@@href}%
\providecommand \@@href[1]{\endgroup#1\@@endlink}%
\providecommand \@sanitize@url [0]{\catcode `\\12\catcode `\$12\catcode
  `\&12\catcode `\#12\catcode `\^12\catcode `\_12\catcode `\%12\relax}%
\providecommand \@@startlink[1]{}%
\providecommand \@@endlink[0]{}%
\providecommand \url  [0]{\begingroup\@sanitize@url \@url }%
\providecommand \@url [1]{\endgroup\@href {#1}{\urlprefix }}%
\providecommand \urlprefix  [0]{URL }%
\providecommand \Eprint [0]{\href }%
\providecommand \doibase [0]{http://dx.doi.org/}%
\providecommand \selectlanguage [0]{\@gobble}%
\providecommand \bibinfo  [0]{\@secondoftwo}%
\providecommand \bibfield  [0]{\@secondoftwo}%
\providecommand \translation [1]{[#1]}%
\providecommand \BibitemOpen [0]{}%
\providecommand \bibitemStop [0]{}%
\providecommand \bibitemNoStop [0]{.\EOS\space}%
\providecommand \EOS [0]{\spacefactor3000\relax}%
\providecommand \BibitemShut  [1]{\csname bibitem#1\endcsname}%
\let\auto@bib@innerbib\@empty
\bibitem [{\citenamefont {Grissonnanche}\ \emph {et~al.}(2019)\citenamefont
  {Grissonnanche}, \citenamefont {Legros}, \citenamefont {Badoux},
  \citenamefont {Lefran{\c{c}}ois}, \citenamefont {Zatko}, \citenamefont
  {Lizaire}, \citenamefont {Lalibert{\'e}}, \citenamefont {Gourgout},
  \citenamefont {Zhou}, \citenamefont {Pyon} \emph
  {et~al.}}]{grissonnanche2019giant}%
  \BibitemOpen
  \bibfield  {author} {\bibinfo {author} {\bibfnamefont {G.}~\bibnamefont
  {Grissonnanche}}, \bibinfo {author} {\bibfnamefont {A.}~\bibnamefont
  {Legros}}, \bibinfo {author} {\bibfnamefont {S.}~\bibnamefont {Badoux}},
  \bibinfo {author} {\bibfnamefont {E.}~\bibnamefont {Lefran{\c{c}}ois}},
  \bibinfo {author} {\bibfnamefont {V.}~\bibnamefont {Zatko}}, \bibinfo
  {author} {\bibfnamefont {M.}~\bibnamefont {Lizaire}}, \bibinfo {author}
  {\bibfnamefont {F.}~\bibnamefont {Lalibert{\'e}}}, \bibinfo {author}
  {\bibfnamefont {A.}~\bibnamefont {Gourgout}}, \bibinfo {author}
  {\bibfnamefont {J.-S.}\ \bibnamefont {Zhou}}, \bibinfo {author}
  {\bibfnamefont {S.}~\bibnamefont {Pyon}},  \emph {et~al.},\ }\href@noop {}
  {\bibfield  {journal} {\bibinfo  {journal} {Nature}\ }\textbf {\bibinfo
  {volume} {571}},\ \bibinfo {pages} {376} (\bibinfo {year}
  {2019})}\BibitemShut {NoStop}%
\bibitem [{\citenamefont {Samajdar}\ \emph
  {et~al.}(2019{\natexlab{a}})\citenamefont {Samajdar}, \citenamefont
  {Chatterjee}, \citenamefont {Sachdev},\ and\ \citenamefont
  {Scheurer}}]{samajdar2019thermal}%
  \BibitemOpen
  \bibfield  {author} {\bibinfo {author} {\bibfnamefont {R.}~\bibnamefont
  {Samajdar}}, \bibinfo {author} {\bibfnamefont {S.}~\bibnamefont
  {Chatterjee}}, \bibinfo {author} {\bibfnamefont {S.}~\bibnamefont {Sachdev}},
  \ and\ \bibinfo {author} {\bibfnamefont {M.~S.}\ \bibnamefont {Scheurer}},\
  }\href@noop {} {\bibfield  {journal} {\bibinfo  {journal} {Physical Review
  B}\ }\textbf {\bibinfo {volume} {99}},\ \bibinfo {pages} {165126} (\bibinfo
  {year} {2019}{\natexlab{a}})}\BibitemShut {NoStop}%
\bibitem [{\citenamefont {Samajdar}\ \emph
  {et~al.}(2019{\natexlab{b}})\citenamefont {Samajdar}, \citenamefont
  {Scheurer}, \citenamefont {Chatterjee}, \citenamefont {Guo}, \citenamefont
  {Xu},\ and\ \citenamefont {Sachdev}}]{samajdar2019enhanced}%
  \BibitemOpen
  \bibfield  {author} {\bibinfo {author} {\bibfnamefont {R.}~\bibnamefont
  {Samajdar}}, \bibinfo {author} {\bibfnamefont {M.~S.}\ \bibnamefont
  {Scheurer}}, \bibinfo {author} {\bibfnamefont {S.}~\bibnamefont
  {Chatterjee}}, \bibinfo {author} {\bibfnamefont {H.}~\bibnamefont {Guo}},
  \bibinfo {author} {\bibfnamefont {C.}~\bibnamefont {Xu}}, \ and\ \bibinfo
  {author} {\bibfnamefont {S.}~\bibnamefont {Sachdev}},\ }\href@noop {}
  {\bibfield  {journal} {\bibinfo  {journal} {Nature Physics}\ }\textbf
  {\bibinfo {volume} {15}},\ \bibinfo {pages} {1290} (\bibinfo {year}
  {2019}{\natexlab{b}})}\BibitemShut {NoStop}%
\bibitem [{\citenamefont {Han}\ \emph {et~al.}(2019)\citenamefont {Han},
  \citenamefont {Park},\ and\ \citenamefont {Lee}}]{han2019consideration}%
  \BibitemOpen
  \bibfield  {author} {\bibinfo {author} {\bibfnamefont {J.~H.}\ \bibnamefont
  {Han}}, \bibinfo {author} {\bibfnamefont {J.-H.}\ \bibnamefont {Park}}, \
  and\ \bibinfo {author} {\bibfnamefont {P.~A.}\ \bibnamefont {Lee}},\
  }\href@noop {} {\bibfield  {journal} {\bibinfo  {journal} {Physical Review
  B}\ }\textbf {\bibinfo {volume} {99}},\ \bibinfo {pages} {205157} (\bibinfo
  {year} {2019})}\BibitemShut {NoStop}%
\bibitem [{\citenamefont {Li}\ and\ \citenamefont {Lee}(2019)}]{li2019thermal}%
  \BibitemOpen
  \bibfield  {author} {\bibinfo {author} {\bibfnamefont {Z.-X.}\ \bibnamefont
  {Li}}\ and\ \bibinfo {author} {\bibfnamefont {D.-H.}\ \bibnamefont {Lee}},\
  }\href@noop {} {\bibfield  {journal} {\bibinfo  {journal} {arXiv preprint
  arXiv:1905.04248}\ } (\bibinfo {year} {2019})}\BibitemShut {NoStop}%
\bibitem [{\citenamefont {Li}\ \emph {et~al.}(2020)\citenamefont {Li},
  \citenamefont {Fauqu{\'e}}, \citenamefont {Zhu},\ and\ \citenamefont
  {Behnia}}]{li2020phonon}%
  \BibitemOpen
  \bibfield  {author} {\bibinfo {author} {\bibfnamefont {X.}~\bibnamefont
  {Li}}, \bibinfo {author} {\bibfnamefont {B.}~\bibnamefont {Fauqu{\'e}}},
  \bibinfo {author} {\bibfnamefont {Z.}~\bibnamefont {Zhu}}, \ and\ \bibinfo
  {author} {\bibfnamefont {K.}~\bibnamefont {Behnia}},\ }\href@noop {}
  {\bibfield  {journal} {\bibinfo  {journal} {Physical Review Letters}\
  }\textbf {\bibinfo {volume} {124}},\ \bibinfo {pages} {105901} (\bibinfo
  {year} {2020})}\BibitemShut {NoStop}%
\bibitem [{\citenamefont {Beck}\ \emph {et~al.}(1974)\citenamefont {Beck},
  \citenamefont {Meier},\ and\ \citenamefont {Thellung}}]{beck1974phonon}%
  \BibitemOpen
  \bibfield  {author} {\bibinfo {author} {\bibfnamefont {H.}~\bibnamefont
  {Beck}}, \bibinfo {author} {\bibfnamefont {P.}~\bibnamefont {Meier}}, \ and\
  \bibinfo {author} {\bibfnamefont {A.}~\bibnamefont {Thellung}},\ }\href@noop
  {} {\bibfield  {journal} {\bibinfo  {journal} {physica status solidi (a)}\
  }\textbf {\bibinfo {volume} {24}},\ \bibinfo {pages} {11} (\bibinfo {year}
  {1974})}\BibitemShut {NoStop}%
\bibitem [{Note1()}]{Note1}%
  \BibitemOpen
  \bibinfo {note} {Here $\kappa _H$ is defined as the thermal Hall conductance
  per unit thickness in the $\protect \mathaccentV {hat}05E{\protect \mathbf
  {z}}$ direction. For $\kappa _H$, in contrast with the electronic $\sigma
  _H$, the macroscopically averaged value and the bulk local value are
  drastically different, as there is a substantial cancellation between the
  heat current carried in the bulk and near the edge \cite
  {cooper1997thermoelectric, qin2011energy, qin2012berry}.}\BibitemShut {Stop}%
\bibitem [{\citenamefont {Rowley}\ \emph {et~al.}(2014)\citenamefont {Rowley},
  \citenamefont {Spalek}, \citenamefont {Smith}, \citenamefont {Dean},
  \citenamefont {Itoh}, \citenamefont {Scott}, \citenamefont {Lonzarich},\ and\
  \citenamefont {Saxena}}]{rowley2014ferroelectric}%
  \BibitemOpen
  \bibfield  {author} {\bibinfo {author} {\bibfnamefont {S.}~\bibnamefont
  {Rowley}}, \bibinfo {author} {\bibfnamefont {L.}~\bibnamefont {Spalek}},
  \bibinfo {author} {\bibfnamefont {R.}~\bibnamefont {Smith}}, \bibinfo
  {author} {\bibfnamefont {M.}~\bibnamefont {Dean}}, \bibinfo {author}
  {\bibfnamefont {M.}~\bibnamefont {Itoh}}, \bibinfo {author} {\bibfnamefont
  {J.}~\bibnamefont {Scott}}, \bibinfo {author} {\bibfnamefont
  {G.}~\bibnamefont {Lonzarich}}, \ and\ \bibinfo {author} {\bibfnamefont
  {S.}~\bibnamefont {Saxena}},\ }\href@noop {} {\bibfield  {journal} {\bibinfo
  {journal} {Nature Physics}\ }\textbf {\bibinfo {volume} {10}},\ \bibinfo
  {pages} {367} (\bibinfo {year} {2014})}\BibitemShut {NoStop}%
\bibitem [{\citenamefont {Zubko}\ \emph {et~al.}(2013)\citenamefont {Zubko},
  \citenamefont {Catalan},\ and\ \citenamefont
  {Tagantsev}}]{zubko2013flexoelectric}%
  \BibitemOpen
  \bibfield  {author} {\bibinfo {author} {\bibfnamefont {P.}~\bibnamefont
  {Zubko}}, \bibinfo {author} {\bibfnamefont {G.}~\bibnamefont {Catalan}}, \
  and\ \bibinfo {author} {\bibfnamefont {A.~K.}\ \bibnamefont {Tagantsev}},\
  }\href@noop {} {\bibfield  {journal} {\bibinfo  {journal} {Annual Review of
  Materials Research}\ }\textbf {\bibinfo {volume} {43}} (\bibinfo {year}
  {2013})}\BibitemShut {NoStop}%
\bibitem [{\citenamefont {Qin}\ \emph {et~al.}(2012)\citenamefont {Qin},
  \citenamefont {Zhou},\ and\ \citenamefont {Shi}}]{qin2012berry}%
  \BibitemOpen
  \bibfield  {author} {\bibinfo {author} {\bibfnamefont {T.}~\bibnamefont
  {Qin}}, \bibinfo {author} {\bibfnamefont {J.}~\bibnamefont {Zhou}}, \ and\
  \bibinfo {author} {\bibfnamefont {J.}~\bibnamefont {Shi}},\ }\href@noop {}
  {\bibfield  {journal} {\bibinfo  {journal} {Physical Review B}\ }\textbf
  {\bibinfo {volume} {86}},\ \bibinfo {pages} {104305} (\bibinfo {year}
  {2012})}\BibitemShut {NoStop}%
\bibitem [{\citenamefont {Lyddane}\ \emph {et~al.}(1941)\citenamefont
  {Lyddane}, \citenamefont {Sachs},\ and\ \citenamefont
  {Teller}}]{lyddane1941}%
  \BibitemOpen
  \bibfield  {author} {\bibinfo {author} {\bibfnamefont {R.~H.}\ \bibnamefont
  {Lyddane}}, \bibinfo {author} {\bibfnamefont {R.~G.}\ \bibnamefont {Sachs}},
  \ and\ \bibinfo {author} {\bibfnamefont {E.}~\bibnamefont {Teller}},\ }\href
  {\doibase 10.1103/PhysRev.59.673} {\bibfield  {journal} {\bibinfo  {journal}
  {Phys. Rev.}\ }\textbf {\bibinfo {volume} {59}},\ \bibinfo {pages} {673}
  (\bibinfo {year} {1941})}\BibitemShut {NoStop}%
\bibitem [{\citenamefont {Ruhman}\ and\ \citenamefont
  {Lee}(2016)}]{ruhman2016}%
  \BibitemOpen
  \bibfield  {author} {\bibinfo {author} {\bibfnamefont {J.}~\bibnamefont
  {Ruhman}}\ and\ \bibinfo {author} {\bibfnamefont {P.~A.}\ \bibnamefont
  {Lee}},\ }\href {\doibase 10.1103/PhysRevB.94.224515} {\bibfield  {journal}
  {\bibinfo  {journal} {Phys. Rev. B}\ }\textbf {\bibinfo {volume} {94}},\
  \bibinfo {pages} {224515} (\bibinfo {year} {2016})}\BibitemShut {NoStop}%
\bibitem [{\citenamefont {Dec}\ \emph {et~al.}(1999)\citenamefont {Dec},
  \citenamefont {Kleemann},\ and\ \citenamefont {Westwanski}}]{dec1999scaling}%
  \BibitemOpen
  \bibfield  {author} {\bibinfo {author} {\bibfnamefont {J.}~\bibnamefont
  {Dec}}, \bibinfo {author} {\bibfnamefont {W.}~\bibnamefont {Kleemann}}, \
  and\ \bibinfo {author} {\bibfnamefont {B.}~\bibnamefont {Westwanski}},\
  }\href@noop {} {\bibfield  {journal} {\bibinfo  {journal} {Journal of
  Physics: Condensed Matter}\ }\textbf {\bibinfo {volume} {11}},\ \bibinfo
  {pages} {L379} (\bibinfo {year} {1999})}\BibitemShut {NoStop}%
\bibitem [{\citenamefont {Smit}(1955)}]{smit1955spontaneous}%
  \BibitemOpen
  \bibfield  {author} {\bibinfo {author} {\bibfnamefont {J.}~\bibnamefont
  {Smit}},\ }\href@noop {} {\bibfield  {journal} {\bibinfo  {journal}
  {Physica}\ }\textbf {\bibinfo {volume} {21}},\ \bibinfo {pages} {877}
  (\bibinfo {year} {1955})}\BibitemShut {NoStop}%
\bibitem [{\citenamefont {Smit}(1958)}]{smit1958spontaneous}%
  \BibitemOpen
  \bibfield  {author} {\bibinfo {author} {\bibfnamefont {J.}~\bibnamefont
  {Smit}},\ }\href@noop {} {\bibfield  {journal} {\bibinfo  {journal}
  {Physica}\ }\textbf {\bibinfo {volume} {24}},\ \bibinfo {pages} {39}
  (\bibinfo {year} {1958})}\BibitemShut {NoStop}%
\bibitem [{\citenamefont {Majumdar}\ and\ \citenamefont
  {Berger}(1973)}]{majumdar1973hall}%
  \BibitemOpen
  \bibfield  {author} {\bibinfo {author} {\bibfnamefont {A.}~\bibnamefont
  {Majumdar}}\ and\ \bibinfo {author} {\bibfnamefont {L.}~\bibnamefont
  {Berger}},\ }\href@noop {} {\bibfield  {journal} {\bibinfo  {journal}
  {Physical Review B}\ }\textbf {\bibinfo {volume} {7}},\ \bibinfo {pages}
  {4203} (\bibinfo {year} {1973})}\BibitemShut {NoStop}%
\bibitem [{\citenamefont {Nagaosa}\ \emph {et~al.}(2010)\citenamefont
  {Nagaosa}, \citenamefont {Sinova}, \citenamefont {Onoda}, \citenamefont
  {MacDonald},\ and\ \citenamefont {Ong}}]{nagaosa2010anomalous}%
  \BibitemOpen
  \bibfield  {author} {\bibinfo {author} {\bibfnamefont {N.}~\bibnamefont
  {Nagaosa}}, \bibinfo {author} {\bibfnamefont {J.}~\bibnamefont {Sinova}},
  \bibinfo {author} {\bibfnamefont {S.}~\bibnamefont {Onoda}}, \bibinfo
  {author} {\bibfnamefont {A.~H.}\ \bibnamefont {MacDonald}}, \ and\ \bibinfo
  {author} {\bibfnamefont {N.~P.}\ \bibnamefont {Ong}},\ }\href@noop {}
  {\bibfield  {journal} {\bibinfo  {journal} {Reviews of modern physics}\
  }\textbf {\bibinfo {volume} {82}},\ \bibinfo {pages} {1539} (\bibinfo {year}
  {2010})}\BibitemShut {NoStop}%
\bibitem [{Note2()}]{Note2}%
  \BibitemOpen
  \bibinfo {note} {If $\protect \overline {\Delta H} \not =0$ in electron
  scattering (e.g. when disorders are mostly repulsive \cite
  {ishizuka2017noncommutative}), then the leading Berry curvature induced skew
  scattering occurs at $\protect \overline {(\Delta H)^3} \sim 1/\ell ^{3/2}$,
  hence its contribution to $\sigma _H$ is $1/\ell ^{1/2}$ smaller than $\sigma
  _L$.}\BibitemShut {Stop}%
\bibitem [{\citenamefont {Casimir}(1938)}]{casimir1938note}%
  \BibitemOpen
  \bibfield  {author} {\bibinfo {author} {\bibfnamefont {H.}~\bibnamefont
  {Casimir}},\ }\href@noop {} {\bibfield  {journal} {\bibinfo  {journal}
  {Physica}\ }\textbf {\bibinfo {volume} {5}},\ \bibinfo {pages} {495}
  (\bibinfo {year} {1938})}\BibitemShut {NoStop}%
\bibitem [{\citenamefont {Guyer}\ and\ \citenamefont
  {Krumhansl}(1966{\natexlab{a}})}]{guyer1966solution}%
  \BibitemOpen
  \bibfield  {author} {\bibinfo {author} {\bibfnamefont {R.~A.}\ \bibnamefont
  {Guyer}}\ and\ \bibinfo {author} {\bibfnamefont {J.}~\bibnamefont
  {Krumhansl}},\ }\href@noop {} {\bibfield  {journal} {\bibinfo  {journal}
  {Physical Review}\ }\textbf {\bibinfo {volume} {148}},\ \bibinfo {pages}
  {766} (\bibinfo {year} {1966}{\natexlab{a}})}\BibitemShut {NoStop}%
\bibitem [{\citenamefont {Guyer}\ and\ \citenamefont
  {Krumhansl}(1966{\natexlab{b}})}]{guyer1966thermal}%
  \BibitemOpen
  \bibfield  {author} {\bibinfo {author} {\bibfnamefont {R.}~\bibnamefont
  {Guyer}}\ and\ \bibinfo {author} {\bibfnamefont {J.}~\bibnamefont
  {Krumhansl}},\ }\href@noop {} {\bibfield  {journal} {\bibinfo  {journal}
  {Physical Review}\ }\textbf {\bibinfo {volume} {148}},\ \bibinfo {pages}
  {778} (\bibinfo {year} {1966}{\natexlab{b}})}\BibitemShut {NoStop}%
\bibitem [{\citenamefont {Martelli}\ \emph {et~al.}(2018)\citenamefont
  {Martelli}, \citenamefont {Jim{\'e}nez}, \citenamefont {Continentino},
  \citenamefont {Baggio-Saitovitch},\ and\ \citenamefont
  {Behnia}}]{martelli2018thermal}%
  \BibitemOpen
  \bibfield  {author} {\bibinfo {author} {\bibfnamefont {V.}~\bibnamefont
  {Martelli}}, \bibinfo {author} {\bibfnamefont {J.~L.}\ \bibnamefont
  {Jim{\'e}nez}}, \bibinfo {author} {\bibfnamefont {M.}~\bibnamefont
  {Continentino}}, \bibinfo {author} {\bibfnamefont {E.}~\bibnamefont
  {Baggio-Saitovitch}}, \ and\ \bibinfo {author} {\bibfnamefont
  {K.}~\bibnamefont {Behnia}},\ }\href@noop {} {\bibfield  {journal} {\bibinfo
  {journal} {Physical review letters}\ }\textbf {\bibinfo {volume} {120}},\
  \bibinfo {pages} {125901} (\bibinfo {year} {2018})}\BibitemShut {NoStop}%
\bibitem [{Note3()}]{Note3}%
  \BibitemOpen
  \bibinfo {note} {Between $10-20\protect \mathrm {K}$ the phenomenon of
  Poiseuille flow \cite {guyer1966solution, guyer1966thermal} is observed \cite
  {martelli2018thermal}, where $\kappa _L$ increases faster than $T^3$ with
  $T$. This is because the onset of phonon-phonon interactions decreases the
  true diffusion velocity compared to the ballistic phonon velocity, and hence
  increases the relaxation time. Similar increase is observed in $\kappa _H$ in
  the same temperature range \cite {li2020phonon}, in support of our
  interpretation of $\kappa _H$ as an extrinsic effect.}\BibitemShut {Stop}%
\bibitem [{\citenamefont {Buckley}\ \emph {et~al.}(1999)\citenamefont
  {Buckley}, \citenamefont {Rivera},\ and\ \citenamefont
  {Salje}}]{buckley1999twin}%
  \BibitemOpen
  \bibfield  {author} {\bibinfo {author} {\bibfnamefont {A.}~\bibnamefont
  {Buckley}}, \bibinfo {author} {\bibfnamefont {J.-P.}\ \bibnamefont {Rivera}},
  \ and\ \bibinfo {author} {\bibfnamefont {E.~K.}\ \bibnamefont {Salje}},\
  }\href@noop {} {\bibfield  {journal} {\bibinfo  {journal} {Journal of applied
  physics}\ }\textbf {\bibinfo {volume} {86}},\ \bibinfo {pages} {1653}
  (\bibinfo {year} {1999})}\BibitemShut {NoStop}%
\bibitem [{\citenamefont {Scott}\ \emph {et~al.}(2012)\citenamefont {Scott},
  \citenamefont {Salje},\ and\ \citenamefont {Carpenter}}]{scott2012domain}%
  \BibitemOpen
  \bibfield  {author} {\bibinfo {author} {\bibfnamefont {J.}~\bibnamefont
  {Scott}}, \bibinfo {author} {\bibfnamefont {E.}~\bibnamefont {Salje}}, \ and\
  \bibinfo {author} {\bibfnamefont {M.}~\bibnamefont {Carpenter}},\ }\href@noop
  {} {\bibfield  {journal} {\bibinfo  {journal} {Physical Review Letters}\
  }\textbf {\bibinfo {volume} {109}},\ \bibinfo {pages} {187601} (\bibinfo
  {year} {2012})}\BibitemShut {NoStop}%
\bibitem [{\citenamefont {Kogan}(1964)}]{kogan1964piezoelectric}%
  \BibitemOpen
  \bibfield  {author} {\bibinfo {author} {\bibfnamefont {S.~M.}\ \bibnamefont
  {Kogan}},\ }\href@noop {} {\bibfield  {journal} {\bibinfo  {journal} {Soviet
  Physics-Solid State}\ }\textbf {\bibinfo {volume} {5}},\ \bibinfo {pages}
  {2069} (\bibinfo {year} {1964})}\BibitemShut {NoStop}%
\bibitem [{\citenamefont {Zubko}\ \emph {et~al.}(2007)\citenamefont {Zubko},
  \citenamefont {Catalan}, \citenamefont {Buckley}, \citenamefont {Welche},\
  and\ \citenamefont {Scott}}]{zubko2007strain}%
  \BibitemOpen
  \bibfield  {author} {\bibinfo {author} {\bibfnamefont {P.}~\bibnamefont
  {Zubko}}, \bibinfo {author} {\bibfnamefont {G.}~\bibnamefont {Catalan}},
  \bibinfo {author} {\bibfnamefont {A.}~\bibnamefont {Buckley}}, \bibinfo
  {author} {\bibfnamefont {P.}~\bibnamefont {Welche}}, \ and\ \bibinfo {author}
  {\bibfnamefont {J.}~\bibnamefont {Scott}},\ }\href@noop {} {\bibfield
  {journal} {\bibinfo  {journal} {Physical Review Letters}\ }\textbf {\bibinfo
  {volume} {99}},\ \bibinfo {pages} {167601} (\bibinfo {year}
  {2007})}\BibitemShut {NoStop}%
\bibitem [{\citenamefont {Zubko}\ \emph {et~al.}(2008)\citenamefont {Zubko},
  \citenamefont {Catalan}, \citenamefont {Buckley}, \citenamefont {Welche},\
  and\ \citenamefont {Scott}}]{zubko2008erratum}%
  \BibitemOpen
  \bibfield  {author} {\bibinfo {author} {\bibfnamefont {P.}~\bibnamefont
  {Zubko}}, \bibinfo {author} {\bibfnamefont {G.}~\bibnamefont {Catalan}},
  \bibinfo {author} {\bibfnamefont {A.}~\bibnamefont {Buckley}}, \bibinfo
  {author} {\bibfnamefont {P.}~\bibnamefont {Welche}}, \ and\ \bibinfo {author}
  {\bibfnamefont {J.}~\bibnamefont {Scott}},\ }\href@noop {} {\bibfield
  {journal} {\bibinfo  {journal} {Physical Review Letters}\ }\textbf {\bibinfo
  {volume} {100}},\ \bibinfo {pages} {199906(E)} (\bibinfo {year}
  {2008})}\BibitemShut {NoStop}%
\bibitem [{\citenamefont {Hong}\ \emph {et~al.}(2010)\citenamefont {Hong},
  \citenamefont {Catalan}, \citenamefont {Scott},\ and\ \citenamefont
  {Artacho}}]{hong2010flexoelectricity}%
  \BibitemOpen
  \bibfield  {author} {\bibinfo {author} {\bibfnamefont {J.}~\bibnamefont
  {Hong}}, \bibinfo {author} {\bibfnamefont {G.}~\bibnamefont {Catalan}},
  \bibinfo {author} {\bibfnamefont {J.}~\bibnamefont {Scott}}, \ and\ \bibinfo
  {author} {\bibfnamefont {E.}~\bibnamefont {Artacho}},\ }\href@noop {}
  {\bibfield  {journal} {\bibinfo  {journal} {Journal of Physics: Condensed
  Matter}\ }\textbf {\bibinfo {volume} {22}},\ \bibinfo {pages} {112201}
  (\bibinfo {year} {2010})}\BibitemShut {NoStop}%
\bibitem [{\citenamefont {Scott}\ and\ \citenamefont
  {Ledbetter}(1997)}]{scott1997interpretation}%
  \BibitemOpen
  \bibfield  {author} {\bibinfo {author} {\bibfnamefont {J.}~\bibnamefont
  {Scott}}\ and\ \bibinfo {author} {\bibfnamefont {H.}~\bibnamefont
  {Ledbetter}},\ }\href@noop {} {\bibfield  {journal} {\bibinfo  {journal}
  {Zeitschrift f{\"u}r Physik B Condensed Matter}\ }\textbf {\bibinfo {volume}
  {104}},\ \bibinfo {pages} {635} (\bibinfo {year} {1997})}\BibitemShut
  {NoStop}%
\bibitem [{Note4()}]{Note4}%
  \BibitemOpen
  \bibinfo {note} {The measured values of $\kappa _H$ appear to vary by of
  order a factor of two for different samples, and even for the same sample if
  a thermal cycle is performed across the tetragonal transition temperature at
  $105\protect \mathrm {K}$. This is taken as evidence that $\kappa _H$ is due
  extrinsically to twin boundaries \cite {li2020phonon}, although why the
  dependence on thermal history is larger in $\kappa _H$ than in $\kappa _L$
  remains unclear without a modeling of the scattering process.}\BibitemShut
  {Stop}%
\bibitem [{Note5()}]{Note5}%
  \BibitemOpen
  \bibinfo {note} {We have used the bulk value of $\chi \epsilon _0 F$. Near
  the twin boundaries where the scatterings occur, the local value is likely
  different (probably enhanced) due to the localized dipoles and structures
  \cite {zubko2007strain, scott2012domain, morozovska2012impact}.}\BibitemShut
  {Stop}%
\bibitem [{\citenamefont {Salce}\ \emph {et~al.}(1994)\citenamefont {Salce},
  \citenamefont {Gravil},\ and\ \citenamefont {Boatner}}]{salce1994disorder}%
  \BibitemOpen
  \bibfield  {author} {\bibinfo {author} {\bibfnamefont {B.}~\bibnamefont
  {Salce}}, \bibinfo {author} {\bibfnamefont {J.}~\bibnamefont {Gravil}}, \
  and\ \bibinfo {author} {\bibfnamefont {L.}~\bibnamefont {Boatner}},\
  }\href@noop {} {\bibfield  {journal} {\bibinfo  {journal} {Journal of
  Physics: Condensed Matter}\ }\textbf {\bibinfo {volume} {6}},\ \bibinfo
  {pages} {4077} (\bibinfo {year} {1994})}\BibitemShut {NoStop}%
\bibitem [{\citenamefont {Chen}\ \emph {et~al.}(1991)\citenamefont {Chen},
  \citenamefont {Birgeneau}, \citenamefont {Kastner}, \citenamefont {Preyer},\
  and\ \citenamefont {Thio}}]{chen1991frequency}%
  \BibitemOpen
  \bibfield  {author} {\bibinfo {author} {\bibfnamefont {C.}~\bibnamefont
  {Chen}}, \bibinfo {author} {\bibfnamefont {R.}~\bibnamefont {Birgeneau}},
  \bibinfo {author} {\bibfnamefont {M.}~\bibnamefont {Kastner}}, \bibinfo
  {author} {\bibfnamefont {N.}~\bibnamefont {Preyer}}, \ and\ \bibinfo {author}
  {\bibfnamefont {T.}~\bibnamefont {Thio}},\ }\href@noop {} {\bibfield
  {journal} {\bibinfo  {journal} {Physical Review B}\ }\textbf {\bibinfo
  {volume} {43}},\ \bibinfo {pages} {392} (\bibinfo {year} {1991})}\BibitemShut
  {NoStop}%
\bibitem [{\citenamefont {Frachet}\ \emph {et~al.}(2019)\citenamefont
  {Frachet}, \citenamefont {Vinograd}, \citenamefont {Zhou}, \citenamefont
  {Benhabib}, \citenamefont {Wu}, \citenamefont {Mayaffre}, \citenamefont
  {Kr\"{a}mer}, \citenamefont {Ramakrishna}, \citenamefont {Reyes},
  \citenamefont {Debray}, \citenamefont {Kurosawa}, \citenamefont {Momono},
  \citenamefont {Oda}, \citenamefont {Komiya}, \citenamefont {Ono},
  \citenamefont {Horio}, \citenamefont {Chang}, \citenamefont {Proust},
  \citenamefont {LeBoeuf},\ and\ \citenamefont {Julien}}]{frachet2019hidden}%
  \BibitemOpen
  \bibfield  {author} {\bibinfo {author} {\bibfnamefont {M.}~\bibnamefont
  {Frachet}}, \bibinfo {author} {\bibfnamefont {I.}~\bibnamefont {Vinograd}},
  \bibinfo {author} {\bibfnamefont {R.}~\bibnamefont {Zhou}}, \bibinfo {author}
  {\bibfnamefont {S.}~\bibnamefont {Benhabib}}, \bibinfo {author}
  {\bibfnamefont {S.}~\bibnamefont {Wu}}, \bibinfo {author} {\bibfnamefont
  {H.}~\bibnamefont {Mayaffre}}, \bibinfo {author} {\bibfnamefont
  {S.}~\bibnamefont {Kr\"{a}mer}}, \bibinfo {author} {\bibfnamefont {S.~K.}\
  \bibnamefont {Ramakrishna}}, \bibinfo {author} {\bibfnamefont
  {A.}~\bibnamefont {Reyes}}, \bibinfo {author} {\bibfnamefont
  {J.}~\bibnamefont {Debray}}, \bibinfo {author} {\bibfnamefont
  {T.}~\bibnamefont {Kurosawa}}, \bibinfo {author} {\bibfnamefont
  {N.}~\bibnamefont {Momono}}, \bibinfo {author} {\bibfnamefont
  {M.}~\bibnamefont {Oda}}, \bibinfo {author} {\bibfnamefont {S.}~\bibnamefont
  {Komiya}}, \bibinfo {author} {\bibfnamefont {S.}~\bibnamefont {Ono}},
  \bibinfo {author} {\bibfnamefont {M.}~\bibnamefont {Horio}}, \bibinfo
  {author} {\bibfnamefont {J.}~\bibnamefont {Chang}}, \bibinfo {author}
  {\bibfnamefont {C.}~\bibnamefont {Proust}}, \bibinfo {author} {\bibfnamefont
  {D.}~\bibnamefont {LeBoeuf}}, \ and\ \bibinfo {author} {\bibfnamefont
  {M.-H.}\ \bibnamefont {Julien}},\ }\href@noop {} {\  (\bibinfo {year}
  {2019})},\ \Eprint {http://arxiv.org/abs/1909.10258} {arXiv:1909.10258
  [cond-mat.supr-con]} \BibitemShut {NoStop}%
\bibitem [{\citenamefont {Grissonnanche}\ \emph {et~al.}(2020)\citenamefont
  {Grissonnanche}, \citenamefont {Th{\'e}riault}, \citenamefont {Gourgout},
  \citenamefont {Boulanger}, \citenamefont {Lefran{\c{c}}ois}, \citenamefont
  {Ataei}, \citenamefont {Lalibert{\'e}}, \citenamefont {Dion}, \citenamefont
  {Zhou}, \citenamefont {Pyon} \emph {et~al.}}]{newLouis}%
  \BibitemOpen
  \bibfield  {author} {\bibinfo {author} {\bibfnamefont {G.}~\bibnamefont
  {Grissonnanche}}, \bibinfo {author} {\bibfnamefont {S.}~\bibnamefont
  {Th{\'e}riault}}, \bibinfo {author} {\bibfnamefont {A.}~\bibnamefont
  {Gourgout}}, \bibinfo {author} {\bibfnamefont {M.-E.}\ \bibnamefont
  {Boulanger}}, \bibinfo {author} {\bibfnamefont {E.}~\bibnamefont
  {Lefran{\c{c}}ois}}, \bibinfo {author} {\bibfnamefont {A.}~\bibnamefont
  {Ataei}}, \bibinfo {author} {\bibfnamefont {F.}~\bibnamefont
  {Lalibert{\'e}}}, \bibinfo {author} {\bibfnamefont {M.}~\bibnamefont {Dion}},
  \bibinfo {author} {\bibfnamefont {J.-S.}\ \bibnamefont {Zhou}}, \bibinfo
  {author} {\bibfnamefont {S.}~\bibnamefont {Pyon}},  \emph {et~al.},\
  }\href@noop {} {\bibfield  {journal} {\bibinfo  {journal} {arXiv preprint
  arXiv:2003.00111}\ } (\bibinfo {year} {2020})}\BibitemShut {NoStop}%
\bibitem [{\citenamefont {Cooper}\ \emph {et~al.}(1997)\citenamefont {Cooper},
  \citenamefont {Halperin},\ and\ \citenamefont
  {Ruzin}}]{cooper1997thermoelectric}%
  \BibitemOpen
  \bibfield  {author} {\bibinfo {author} {\bibfnamefont {N.}~\bibnamefont
  {Cooper}}, \bibinfo {author} {\bibfnamefont {B.}~\bibnamefont {Halperin}}, \
  and\ \bibinfo {author} {\bibfnamefont {I.}~\bibnamefont {Ruzin}},\
  }\href@noop {} {\bibfield  {journal} {\bibinfo  {journal} {Physical Review
  B}\ }\textbf {\bibinfo {volume} {55}},\ \bibinfo {pages} {2344} (\bibinfo
  {year} {1997})}\BibitemShut {NoStop}%
\bibitem [{\citenamefont {Qin}\ \emph {et~al.}(2011)\citenamefont {Qin},
  \citenamefont {Niu},\ and\ \citenamefont {Shi}}]{qin2011energy}%
  \BibitemOpen
  \bibfield  {author} {\bibinfo {author} {\bibfnamefont {T.}~\bibnamefont
  {Qin}}, \bibinfo {author} {\bibfnamefont {Q.}~\bibnamefont {Niu}}, \ and\
  \bibinfo {author} {\bibfnamefont {J.}~\bibnamefont {Shi}},\ }\href@noop {}
  {\bibfield  {journal} {\bibinfo  {journal} {Physical review letters}\
  }\textbf {\bibinfo {volume} {107}},\ \bibinfo {pages} {236601} (\bibinfo
  {year} {2011})}\BibitemShut {NoStop}%
\bibitem [{\citenamefont {Ishizuka}\ and\ \citenamefont
  {Nagaosa}(2017)}]{ishizuka2017noncommutative}%
  \BibitemOpen
  \bibfield  {author} {\bibinfo {author} {\bibfnamefont {H.}~\bibnamefont
  {Ishizuka}}\ and\ \bibinfo {author} {\bibfnamefont {N.}~\bibnamefont
  {Nagaosa}},\ }\href@noop {} {\bibfield  {journal} {\bibinfo  {journal}
  {Physical Review B}\ }\textbf {\bibinfo {volume} {96}},\ \bibinfo {pages}
  {165202} (\bibinfo {year} {2017})}\BibitemShut {NoStop}%
\bibitem [{\citenamefont {Morozovska}\ \emph {et~al.}(2012)\citenamefont
  {Morozovska}, \citenamefont {Eliseev}, \citenamefont {Glinchuk},
  \citenamefont {Chen}, \citenamefont {Kalinin},\ and\ \citenamefont
  {Gopalan}}]{morozovska2012impact}%
  \BibitemOpen
  \bibfield  {author} {\bibinfo {author} {\bibfnamefont {A.}~\bibnamefont
  {Morozovska}}, \bibinfo {author} {\bibfnamefont {E.}~\bibnamefont {Eliseev}},
  \bibinfo {author} {\bibfnamefont {M.~D.}\ \bibnamefont {Glinchuk}}, \bibinfo
  {author} {\bibfnamefont {L.~Q.}\ \bibnamefont {Chen}}, \bibinfo {author}
  {\bibfnamefont {S.}~\bibnamefont {Kalinin}}, \ and\ \bibinfo {author}
  {\bibfnamefont {V.}~\bibnamefont {Gopalan}},\ }\href@noop {} {\bibfield
  {journal} {\bibinfo  {journal} {Ferroelectrics}\ }\textbf {\bibinfo {volume}
  {438}},\ \bibinfo {pages} {32} (\bibinfo {year} {2012})}\BibitemShut
  {NoStop}%
\end{thebibliography}%
\bibliographystyle{apsrev4-1} 

\onecolumngrid
\newpage


\section*{Supplementary material}

# Enhanced thermal Hall effect in nearly ferroelectric insulators: Supplemental Material

Jing-Yuan Chen,[1] Steven A. Kivelson,[2] and Xiao-Qi Sun[2, 3]

[1]*Stanford Institute for Theoretical Physics, Stanford University, Stanford, CA 94305, USA*
[2]*Department of Physics, Stanford University, Stanford, CA 94305, USA*
[3]*Stanford Center for Topological Quantum Physics, Stanford University, Stanford, CA 94305, USA*

In this Supplemental Material we present the details of the analysis of the Berry curvature and intrinsic thermal Hall effect of phonons. First we consider the contribution from acoustic phonons only, reproducing the results of Ref. [1] with some rectification. Then we include the contribution from the lowest optical phonons, which, in materials near ferroelectric quantum criticality, may have a small gap that is comparable to the temperatures in the $\kappa_H$ measurements. We discuss our result mainly in the context of the recently measured thermal Hall conductivity in STO [2]. However, the intrinsic effect is ultimately insignificant in STO – it is our hope that the present results will be more directly relevant to other systems as discussed in the Main Text, which have large flexoelectric couplings but lack of structural defects to host a large extrinsic thermal Hall effect.

We start with a general non-interacting theory of bosons and introduce the general notion of boson Berry curvature. We Fourier transform the real bosonic fields $\phi^i(\mathbf{r})$ (complex bosons can be represented as two real bosons) into $\mathbf{k}$-space, $\phi^i(\mathbf{k}) \equiv \int d^3r \, e^{-i\mathbf{k}\cdot\mathbf{r}/\hbar} \phi(\mathbf{r})$, where $i = 1, \ldots, n$ runs over all boson components (e.g. for acoustic phonons, $i$ runs over the $x, y, z$ directions of displacement, while if we include optical phonons as well, then it also runs over the optical motions). The associated canonical momenta $\pi_j(\mathbf{k})$ are defined by the commutation relations

$$\left[ \begin{pmatrix} \phi^i(\mathbf{k}) \\ \pi_j(\mathbf{k}) \end{pmatrix}, \begin{pmatrix} \phi^{i'}(\mathbf{k}')^\dagger & \pi_{j'}(\mathbf{k}')^\dagger \end{pmatrix} \right] = i\hbar \, (2\pi\hbar)^3 \delta^3(\mathbf{k} - \mathbf{k}') \begin{pmatrix} 0 & \delta^i_{j'} \\ -\delta^{i'}_j & 0 \end{pmatrix}. \tag{1}$$

The reality condition requires $\phi^i(-\mathbf{k}) = \phi^i(\mathbf{k})^\dagger$, and $\pi_j(-\mathbf{k}) = \pi_j(\mathbf{k})^\dagger$. We denote the phase space column vector $(\phi^i(\mathbf{k}) \ \pi_j(\mathbf{k}))^T$ as $\zeta^I(\mathbf{k})$ where the (upper) $I$ runs over both (upper) $i = 1, \ldots, n$ and (lower) $j = 1, \ldots, n$. The commutation relations above can be compactly written as

$$\left[ \zeta^I(\mathbf{k}), \zeta^{I'}(\mathbf{k}')^\dagger \right] = i\hbar \, (2\pi\hbar)^3 \delta^3(\mathbf{k} - \mathbf{k}') \, \mathcal{J}^{II'} \tag{2}$$

(the matrix $\mathcal{J}^{II'}$ is the inverse of the standard symplectic 2-form). A non-interacting Hamiltonian can be written as $H = \int [d^3k/(2\pi\hbar)^3] H(\mathbf{k})$, with $H(\mathbf{k})$ taking the general from

$$H(\mathbf{k}) = \frac{1}{2} \, \zeta^{I'}(\mathbf{k})^\dagger \, \mathcal{H}_{I'I}(\mathbf{k}) \, \zeta^I(\mathbf{k})$$

$$= \frac{1}{2} \begin{pmatrix} \phi^{i'}(\mathbf{k})^\dagger & \pi_{j'}(\mathbf{k})^\dagger \end{pmatrix} \begin{pmatrix} (\mathcal{H}_{\phi\phi})_{i'i}(\mathbf{k}) & (\mathcal{H}_{\phi\pi})_{i'}{}^j(\mathbf{k}) \\ (\mathcal{H}_{\pi\phi})^{j'}{}_i(\mathbf{k}) & (\mathcal{H}_{\pi\pi})^{j'j}(\mathbf{k}) \end{pmatrix} \begin{pmatrix} \phi^i(\mathbf{k}) \\ \pi_j(\mathbf{k}) \end{pmatrix} \tag{3}$$

where $\mathcal{H}_{I'I}(\mathbf{k})$ is Hermitian and positive-definite (for any $\mathbf{k} \neq 0$), and $\mathcal{H}_{I'I}(\mathbf{k}) = \mathcal{H}_{II'}(-\mathbf{k}) = \mathcal{H}^*_{I'I}(-\mathbf{k})$. The evolution of the phase space operators is

$$i\hbar \dot{\zeta}^I(\mathbf{k}) = [\zeta^I(\mathbf{k}), H] = i\hbar \mathcal{J}^{II'} \mathcal{H}_{I'J}(\mathbf{k}) \, \zeta^J(\mathbf{k}). \tag{4}$$

To solve for the dynamics, we tranform to variables $\zeta^I \to M^A_I \zeta^I$ that evolve diagonally as $i\hbar M^A_I \dot{\zeta}^I = \mathcal{E}_A M^A_I \dot{\zeta}^I$ (here $A$ runs from 1 to $2n$). This requires us to diagonalize the (non-Hermitian) matrix $i\hbar \mathcal{J}^{II'} \mathcal{H}_{I'J}(\mathbf{k})$:

$$M^A_I(\mathbf{k}) \, i\hbar \mathcal{J}^{II'} \mathcal{H}_{I'J}(\mathbf{k}) = \mathcal{E}^A{}_{A'}(\mathbf{k}) \, M^{A'}_J(\mathbf{k}) \tag{5}$$

where $\mathcal{E}^A{}_{A'} = \mathcal{E}_A \delta^A_{A'}$ is diagonal. The diagonalization has two important properties. First, we contract the above on the right with the Hermitian conjugate of $M \, i\hbar \mathcal{J}$, and find

$$M^A_I(\mathbf{k}) \, i\hbar \mathcal{J}^{II'} \, \mathcal{H}_{I'J'}(\mathbf{k}) \, i\hbar \mathcal{J}^{J'J} M^B_J(\mathbf{k})^* = \mathcal{E}_A(\mathbf{k}) \, M^A_I(\mathbf{k}) \, i\hbar \mathcal{J}^{IJ} \, M^B_J(\mathbf{k})^*. \tag{6}$$

Since $\mathcal{H}$ is Hermitian and positive-definite, so is the entire left-hand-side. Moreover, since $i\hbar \mathcal{J}$ is Hermitian, so is $M^A_I \, i\hbar \mathcal{J}^{IJ} M^B_J{}^*$ on the right-hand-side. Along with the fact that $\mathcal{E}$ is diagonal, we can conclude that $M^A_I \, i\hbar \mathcal{J}^{IJ} M^B_J{}^*$

must be diagonal (when $\mathcal{E}$ has degeneracy, we can choose a basis so that it is diagonal). Now that the left-hand-side is also diagonal, along with its positive-definiteness, we conclude each eigenvalue $\mathcal{E}_A(\mathbf{k})$ must be real and non-zero, and consequently, we can normalize the eigenvectors so that

$$M^A{}_I(\mathbf{k})\ i\hbar\mathcal{J}^{IJ} M^B{}_J(\mathbf{k})^* = \delta^{AB}\ \text{sgn}\,\mathcal{E}_A(\mathbf{k}). \tag{7}$$

Second, we may take the complex conjugate of Eq.(5) and replace $\mathbf{k}\to-\mathbf{k}$, and use the fact $\mathcal{H}_{I'I}(\mathbf{k})=\mathcal{H}^*_{I'I}(-\mathbf{k})$, to obtain

$$M^A{}_I(-\mathbf{k})^*\ i\hbar\mathcal{J}^{II'}\mathcal{H}_{I'J}(\mathbf{k}) = -\mathcal{E}^A{}_{A'}(-\mathbf{k})^*\ M^{A'}{}_J(-\mathbf{k})^*. \tag{8}$$

Combined with the fact that the eigenvalues are all real and non-zero, this means half of the eigenvalues are positive and half are negative, such that they can be organized as

$$\mathcal{E}^A{}_{A'}(\mathbf{k}) = \begin{pmatrix} E_\alpha(\mathbf{k})\,\delta^\alpha{}_{\alpha'} & 0 \\ 0 & -E_\beta(-\mathbf{k})\,\delta_\beta{}^{\beta'} \end{pmatrix} \tag{9}$$

where (upper) $A$ runs over (upper) $\alpha=1,\ldots,n$ and (lower) $\beta=1,\ldots,n$, and we have denoted the positive eigenvalues as $E_\alpha(\mathbf{k})$, interpreted as the energy eigenvalues of diagonalized modes $\alpha$ at momentum $\mathbf{k}$. Moreover, the associated left eigenvectors can be organized as

$$M^A{}_I(\mathbf{k}) = \begin{pmatrix} M^\alpha{}_I(\mathbf{k}) \\ \delta_{\beta\beta'} M^{\beta'}{}_I(-\mathbf{k})^* \end{pmatrix} = \begin{pmatrix} (M_\phi)^\alpha{}_i(\mathbf{k}) & \delta^{\alpha\alpha'}(M_\pi)_{\alpha'}{}^j(\mathbf{k}) \\ \delta_{\beta\beta'}(M_\phi)^{\beta'}{}_i(-\mathbf{k})^* & (M_\pi)_\beta{}^j(-\mathbf{k})^* \end{pmatrix}. \tag{10}$$

Note that $M$ admits a gauge ambiguity of multiplying on its left a matrix that commutes with $\mathcal{E}$. With the normalization as in Eq.(7), the change of variables we motivated are simply the annihilation and creation operators

$$\begin{pmatrix} a^\alpha(\mathbf{k}) \\ a_\beta(-\mathbf{k})^\dagger \end{pmatrix} \equiv M^A{}_I(\mathbf{k})\,\zeta^I(\mathbf{k}) = \begin{pmatrix} (M_\phi)^\alpha{}_i(\mathbf{k}) & \delta^{\alpha\alpha'}(M_\pi)_{\alpha'}{}^j(\mathbf{k}) \\ \delta_{\beta\beta'}(M_\phi)^{\beta'}{}_i(-\mathbf{k})^* & (M_\pi)_\beta{}^j(-\mathbf{k})^* \end{pmatrix} \begin{pmatrix} \phi^i(\mathbf{k}) \\ \pi_j(\mathbf{k}) \end{pmatrix} \tag{11}$$

(the reality condition $\zeta^I(\mathbf{k})=\zeta^I(-\mathbf{k})^\dagger$ ensures the definition is consistent under $\mathbf{k}\to-\mathbf{k}$, and $a^\alpha(\mathbf{k})$ and $a^\alpha(-\mathbf{k})$ are unrelated), whose commutation relation is rendered by Eq.(7) as we conjugate Eq.(2) by $M$ and $M^\dagger$. The normalization Eq.(7) also allows us to invert $M$ so to express $\zeta$ in terms of $a, a^\dagger$. Then we can express the Hamiltonian as

$$H(\mathbf{k}) = \frac{1}{2}\sum_\alpha \left(E_\alpha(\mathbf{k})\,a^\dagger_\alpha(\mathbf{k})\,a^\alpha(\mathbf{k}) + E_\alpha(-\mathbf{k})\,a^\alpha(-\mathbf{k})\,a^\dagger_\alpha(\mathbf{k})\right) \tag{12}$$

as desired. This is the general procedure of finding the energy eigenstates.

Having diagonalized the Hamiltonian, the Berry connection and Berry curvature of the diagonalized mode $\alpha$ is defined as

$$\boldsymbol{\mathcal{A}}_\alpha(\mathbf{k}) \equiv i\hbar\,\mathcal{J}^{II'}\,M^\alpha{}_I(\mathbf{k})^*\times\nabla_\mathbf{k} M^\alpha{}_{I'}(\mathbf{k}),$$

$$\boldsymbol{\Omega}_\alpha(\mathbf{k}) \equiv \nabla_\mathbf{k}\times\boldsymbol{\mathcal{A}}_\alpha(\mathbf{k}) = i\hbar\,\mathcal{J}^{II'}\,\nabla_\mathbf{k} M^\alpha{}_I(\mathbf{k})^*\times\nabla_\mathbf{k} M^\alpha{}_{I'}(\mathbf{k}). \tag{13}$$

In Ref. [1] it has been shown that the intrinsic thermal Hall conductivity is given by

$$\kappa_H^{in} = T\sum_\alpha \int_0^\infty d\xi \int \frac{d^3k}{(2\pi\hbar)^3}\,\Omega_\alpha^z(\mathbf{k})\,\Theta\left(\xi-\frac{E_\alpha(\mathbf{k})}{T}\right)\,\xi^2\frac{\partial f_{BE}(\xi)}{\partial\xi} \tag{14}$$

where $\Theta$ is the step function, $f_{BE}(\xi)$ is the Bose-Einstein distribution at energy $E=\xi T$, and $\mathbf{B}=B\hat{\mathbf{z}}$. As mentioned in the Main Text, the $\kappa_H$ is defined as the macroscopically averaged conductivity, not the bulk local conductivity, which can be drastically from the microscopically averaged value, due to a substantial cancellation between the heat currents carried in the bulk and near the edge [1, 3, 4].

After introducing the general concept, we turn to our particular problem of acoustic phonons from the Main Text. The bosonic fields are the acoustic displacements $u^i$ for $i=x,y,z$. When $B=0$, the unperturbed Hamiltonian is

$$H_0(\mathbf{k}) = \frac{1}{2}\begin{pmatrix} u^{i'}(-\mathbf{k}) & \pi_{j'}(-\mathbf{k}) \end{pmatrix} \begin{pmatrix} K_1\delta_{i'i}k^2/\hbar^2 + K_2 k_{i'}k_i/\hbar^2 & 0 \\ 0 & \delta^{j'j}/\rho \end{pmatrix} \begin{pmatrix} u^i(\mathbf{k}) \\ \pi_j(\mathbf{k}) \end{pmatrix} \tag{15}$$

whose $\pi_i$ equation of motion $\dot{u}^i(\mathbf{k}) = \delta^{ij}\pi_j(\mathbf{k})/\rho$ relates $H_0$ to the Lagrangian $L_0$ and vice versa. Thanks to rotational invariance, we can first apply a rotation to diagonalize $H_0$ as decoupled oscillators and then find the respective (unperturbed) annihilation and creation operators:

$$(M_0)^A_I(\mathbf{k}) = \frac{1}{\sqrt{2\hbar}} \begin{pmatrix} \delta^\alpha_{\alpha'} & i\delta^{\alpha\beta'} \\ \delta_{\beta\alpha'} & -i\delta^{\beta'}_\beta \end{pmatrix} \begin{pmatrix} \sqrt{\frac{\rho v_{\alpha'}k}{\hbar}}\,(R_{\hat{\mathbf{k}}})^{\alpha'}{}_i & 0 \\ 0 & \sqrt{\frac{\hbar}{\rho v_{\beta'}k}}\delta_{\beta'\beta''}(R_{\hat{\mathbf{k}}})^{\beta''}{}_{\tilde{j}}\,\delta^{\tilde{j}j} \end{pmatrix} \quad (16)$$

where $k \equiv |\mathbf{k}|$, and $v_{\alpha=1,2} \equiv v_T \equiv \sqrt{K_1/\rho}$ is the transverse sound velocity and $v_{\alpha=3} \equiv v_L \equiv \sqrt{(K_1+K_2)/\rho}$ the longitudinal; the energy eigenvalues are $(E_0)_\alpha = v_\alpha k$. Here $(R_{\hat{\mathbf{k}}})^\alpha{}_i$ is any rotation such that $\hat{z}_\alpha(R_{\hat{\mathbf{k}}})^\alpha{}_i = \pm \hat{k}_i$. For definiteness we choose it to be so that, when $k^z > 0$, $R^T_{\hat{\mathbf{k}}}$ is a rotation that follows the Euler angles $(0, \theta_{\hat{\mathbf{k}}}, \phi_{\hat{\mathbf{k}}})$ from $\hat{z}$ to $\hat{\mathbf{k}}$, while when $k^z < 0$ we use the rotation for $-\hat{\mathbf{k}}$; on the $k^z = 0$ equator the two sides differ by gauge transformation we mentioned before, which does not affect the Berry curvature.

When we turn on the $B$, the Hamiltonian changes from $H_0$ by a replacement of the column vector:

$$\begin{pmatrix} u^i(\mathbf{k}) \\ \pi_j(\mathbf{k}) \end{pmatrix} \longrightarrow \begin{pmatrix} \delta^i_{i'} & 0 \\ \varepsilon_{jkl}\frac{\chi\epsilon_0 B^l}{\hbar^2 c}\left(F_1 k^2 \delta^k_{i'} + F_2 k^k k_{i'}\right) & \delta^j_{j'} \end{pmatrix} \begin{pmatrix} u^{i'}(\mathbf{k}) \\ \pi_{j'}(\mathbf{k}) \end{pmatrix} \quad (17)$$

and likewise for the row vector $\begin{pmatrix} u^{i'}(\mathbf{k})^\dagger & \pi_{j'}(\mathbf{k})^\dagger \end{pmatrix}$. This amounts to a change in the physical displacement momentum $\rho \dot{u}_j(\mathbf{k})$ from the canonical momentum $\pi_j(\mathbf{k})$ to a shifted $\pi_j(\mathbf{k}) + \varepsilon_{jkl}(\chi\epsilon_0 B^l/\hbar^2 c)\left(F_1 k^2 \delta^k_{i'} + F_2 k^k k_{i'}\right) u^{i'}(\mathbf{k})$ due to the $L_B$ in the Lagrangian. We can find the new diagonalization $M^A_I(\mathbf{k})$ in perturbation theory to linear order in $B$. The energy eigenvalues become

$$E_{1,2}(\mathbf{k}) = v_T k \mp \frac{\chi\epsilon_0 F_1 \mathbf{B}\cdot \mathbf{k}\, k}{\rho c \hbar}, \qquad E_3(\mathbf{k}) = v_L k, \quad (18)$$

and the associated Berry curvatures are (the $M^\alpha_I$ components are too cumbersome for presentation) [1]

$$\boldsymbol{\Omega}_{1,2}(\mathbf{k}) = \mp \hbar \frac{\mathbf{k}}{k^3} + \frac{\chi\epsilon_0(2F_1+F_2)\left(\mathbf{B}k^2 + (\mathbf{B}\cdot\mathbf{k})\mathbf{k}\right)}{2\rho c v_T k^3} \frac{\Delta^3 + 3\Delta}{2\Delta(\Delta^2-1)},$$

$$\boldsymbol{\Omega}_3(\mathbf{k}) = \frac{\chi\epsilon_0(2F_1+F_2)\left(\mathbf{B}k^2 + (\mathbf{B}\cdot\mathbf{k})\mathbf{k}\right)}{2\rho c v_T k^3} \frac{-3\Delta^2-1}{\Delta(\Delta^2-1)} \quad (19)$$

where $\Delta \equiv v_L/v_T = \sqrt{1 + K_2/K_1}$ (the divergence at $\Delta = 1$ is an artifact – in that case the modes are nearly degenerate, and upon summation over the modes for $\kappa_H$ the result is finite). The resulting intrinsic thermal Hall conductivity is (which bears some rectification to Ref. [1])

$$\kappa^{in}_H = \frac{4\pi^2}{45}\left(\frac{T}{\hbar v_T}\right)^3 \frac{\chi\epsilon_0 B}{\rho c}\left(F_1 - \frac{2F_1+F_2}{2}\frac{\Delta^4+\Delta^3+4\Delta^2+\Delta+1}{\Delta^3(\Delta+1)}\right). \quad (20)$$

The first term arises from the $\mathbf{k}/k^3$ terms in $\boldsymbol{\Omega}_{1,2}$ and the energy split between these two modes; the second term is due to the order $B$ terms in $\boldsymbol{\Omega}_{1,2,3}$. This detailed expression of $\kappa^{in}_H$ indeed agrees, in order of magnitudes, with our estimate in the Main Text.

For STO, $v_L \approx 7100$m/s $\approx 4.7\times 10^{-2}$ÅeV/$\hbar$ and $v_T \approx 4200$m/s $\approx 2.7\times 10^{-2}$ÅeV/$\hbar$ at low temperatures [5], so that $\Delta \approx 1.7$, and the factor involving $\Delta$ in Eq.(20) is approximately 2. The density $\rho \approx 5100$kg/m$^3 \approx 7.4\times 10^2 \hbar^2/$eVÅ$^5$. The electric susceptibility is $\chi \approx 2\times 10^4$ for temperatures $T \lesssim 10$K [6]. The flexoelectric couplings have been measured [7, 8], but there is an unsettled ambiguity [9] in the interpretation and therefore the component-wise values and signs are presently unclear; it might also be the case that our rotational invariance assumption is not a good approximation. For now we will just assume $F_{1,2}$ in Eq.(20) to take the "normal value" of flexoelectric couplings, $e/4\pi\epsilon_0 a \approx 2\times 10^{-2} e/$Å$\epsilon_0$ (for $a \approx 3.9$Å) [9, 10], which is indeed the order of the values from measurements [7, 8] and numerics [11] (with inconclusive sign). Taking $B/c = 10$T $= 1.5\times 10^{-4}\hbar/$Å$^2$ and $T = 10$K $= 8.6\times 10^{-4}$eV, we find $|\kappa^{in}_H| \approx 4\times 10^{-9}$eV/Å$\hbar \approx 10^{-6}$W/Km. This is $10^{-4}$ times smaller in magnitude than the observed value $\kappa_H \approx -10^{-2}$W/Km (up to a dependence on the sample and thermal history for about a factor of 2) at this temperature and magnetic field [2].

Having considered the intrinsic thermal Hall effect due to acoustic phonons, now we take the lowest optical phonon into consideration. As mentioned in the Main Text, for a nearly ferroelectric insulator the polarization is dominated by a single optical phonon mode $\mathbf{w}$, whose transverse branches become soft, and hence the associated dynamics may be relevant at temperatures comparable to the soft energy gap. More explicitly, we may write down a constitutive relation between the fluctuating dipole order parameter $\mathbf{P}$ and the phonon modes (at $\mathbf{B} = 0$):

$$\mathbf{P} = \rho_e \mathbf{w} + \chi \epsilon_0 F_1 \, \nabla^2 \mathbf{u} + \chi \epsilon_0 F_2 \, \nabla(\nabla \cdot \mathbf{u}) + \ldots \tag{21}$$

where $\mathbf{w}$ is the displacement vector corresponding to the relevant optical modes that carry the dipole moment, the following terms are the flexoelectric effect, and we have omitted terms of higher order in derivatives as well as small contributions from all degrees of freedom with energies higher than $\mathbf{w}$. To take into account the effects of $\mathbf{w}$, we include the action (unperturbed by $\mathbf{B}$ for now)

$$S_0 = \int d^3 \mathbf{r} \left[ \frac{\rho'}{2} \dot{\mathbf{w}}(\mathbf{r})^2 - \frac{K_1'}{2} (\nabla \mathbf{w}(\mathbf{r}))^2 - \frac{K_2'}{2} (\nabla \cdot \mathbf{w}(\mathbf{r}))^2 - \frac{\rho_e^2}{2(\epsilon - \epsilon_\infty)} \mathbf{w}(\mathbf{r})^2 + \ldots \right]$$
$$- \int d^3 \mathbf{r} d^3 \mathbf{r}' \frac{\rho_e^2}{4\pi \epsilon_\infty} \frac{(\nabla \cdot \mathbf{w}(\mathbf{r}))(\nabla' \cdot \mathbf{w}(\mathbf{r}'))}{|\mathbf{r} - \mathbf{r}'|} \, . \tag{22}$$

Note the $\mathbf{w}$ phonons are already defined to be decoupled from the acoustic phonons. The last term is the Coulomb interaction, in which $\epsilon_\infty$, slightly different from $\epsilon_0$, is the dielectric constant at high frequencies, containing renormalization effects from all higher energy degrees of freedom; on the other hand, $\epsilon$ is the static dielectric constant. The diagonalization of the unperturbed Hamiltonian of the $\mathbf{w}$ phonon is similar to that of the $\mathbf{u}$ phonon, with the following replacements

$$\rho \to \rho', \quad K_1 \frac{k^2}{\hbar^2} \to K_1' \frac{k^2}{\hbar^2} + \frac{\rho_e^2}{\epsilon - \epsilon_\infty}, \quad K_2 \frac{k_i k_j}{\hbar^2} \to K_2' \frac{k_i k_j}{\hbar^2} + \frac{\rho_e^2 \hat{k}_i \hat{k}_j}{\epsilon_\infty}, \tag{23}$$

where the non-analyticity near $k = 0$ reflects the long-range interaction. The $\mathbf{w}$ boson energies at $\mathbf{B} = 0$ are found accordingly,

$$(E_0')_1(\mathbf{k}) = (E_0')_2(\mathbf{k}) \equiv E_T'(\mathbf{k}) = \sqrt{\Delta_{\text{TO}}^2 + v_T'^2 k^2}, \quad (E_0')_3(\mathbf{k}) \equiv E_L'(\mathbf{k}) = \sqrt{\Delta_{\text{LO}}^2 + v_L'^2 k^2}, \tag{24}$$

where the $v_T' \equiv \sqrt{K_1'/\rho'}$ and $v_L' \equiv \sqrt{(K_1' + K_2')/\rho'}$ are the characteristic sound speeds of the transverse and longitudinal modes respectively. The transverse and longitudinal optical phonon energies are

$$\Delta_{\text{TO}} \equiv \rho_e \sqrt{\frac{1}{(\epsilon - \epsilon_\infty)\rho'}}, \quad \Delta_{\text{LO}} \equiv \rho_e \sqrt{\frac{1}{(\epsilon - \epsilon_\infty)\rho'} + \frac{1}{\epsilon_\infty \rho'}}, \tag{25}$$

whose ratio is known as the Lyddane-Sachs-Teller relation [12]. As the static dielectric constant $\epsilon$ grows much larger than $\epsilon_\infty$, the transverse optical phonon energy becomes soft, while the longitudinal optical phonon energy remains large.

Now we include the coupling to $\mathbf{B}$. We first consider the coupling between $\mathbf{w}$ and $\dot{\mathbf{u}}$ through the magnetic field,

$$\frac{\mathbf{B}}{c} \cdot (\rho_e \mathbf{w} \times \dot{\mathbf{u}}) + \ldots \tag{26}$$

The leading term is reminiscent of the term $(\mathbf{B}/c) \cdot (\mathbf{P} \times \dot{\mathbf{u}})$, which captures the Lorentz force on a dipole in motion $\dot{\mathbf{u}}$. It is, however, easy to see that at linear order in $\mathbf{B}$ this term makes no contribution to $\kappa_H^{in}$. To show this, we denote the unperturbed annihilation operators associated with the $\mathbf{u}$ and the $\mathbf{w}$ modes as $(a_0)^{\alpha_u} = (M_0)^{\alpha_u}_{I_u} \zeta^{I_u}$ and $(a_0)^{\alpha_w} = (M_0)^{\alpha_w}_{I_w} \zeta^{I_w}$ respectively, where $I_u$ runs over $\mathbf{u}$ and the associated canonical momenta $\boldsymbol{\pi}$, while $I_w$ runs over $\mathbf{w}$ and the associated canonical momenta $\boldsymbol{\varpi}$, and $\alpha_u$, $\alpha_w$ run over the diagonalized annihilation operators of the $\mathbf{u}$ and the $\mathbf{w}$ branches respectively. Denote the linear in $\mathbf{B}$ correction to $M_0$ due to the coupling above as $\delta M$. Consider the Berry connection for the $\alpha_u$ band, which, up to a pure gauge, is proportional to

$$\mathcal{J}^{I_u I_u'} \, (\delta M)^{\alpha_u}_{I_u}(\mathbf{k})^* \, \times \, \nabla_\mathbf{k} (M_0)^{\alpha_u}_{I_u'}(\mathbf{k}) \; + \; \mathcal{J}^{I_w I_w'} \, (\delta M)^{\alpha_u}_{I_w}(\mathbf{k})^* \, \times \, \nabla_\mathbf{k} (M_0)^{\alpha_u}_{I_w'}(\mathbf{k}). \tag{27}$$

However, $\delta M$ has vanishing $^{\alpha_u}_{I_u}$ components because the $\mathbf{B}$ field coupling above is off-diagonal between $\mathbf{u}$ and $\mathbf{w}$; on the other hand, $\nabla_\mathbf{k}(M_0)$ has vanishing $^{\alpha_u}_{I_w'}$ components because in the unperturbed Hamiltonian, $\mathbf{u}$ and $\mathbf{w}$ are separate

at any $\mathbf{k}$. Thus, the $(\mathbf{B}/c) \cdot (\mathbf{w} \times \dot{\mathbf{u}})$ coupling does not contribute, at linear order in $\mathbf{B}$, to the Berry connection of the $\alpha_u$ bands, and likewise the $\alpha_w$ bands, and hence does not contribute to $\kappa_H^{in}$. For the same reason, magnetic coupling terms mixing $\mathbf{w}$ and $\mathbf{u}$ at higher orders in derivatives also make no contribution to $\kappa_H^{in}$ at leading order in $\mathbf{B}$.

Next, we consider the coupling between $\mathbf{w}$ and $\dot{\mathbf{w}}$ through the magnetic field,

$$\frac{\mathbf{B}}{c} \cdot (\rho'_e \mathbf{w} \times \dot{\mathbf{w}}) + \frac{\mathbf{B}}{c} \cdot \chi \epsilon_0 \left( F'_1 \nabla^2 \mathbf{w} + F'_2 \nabla(\nabla \cdot \mathbf{w}) \right) \times \dot{\mathbf{w}}. \tag{28}$$

We are including both a leading and a sub-leading term in derivatives. Before we compute their contributions to $\kappa_H$, we shall explain the origins of these terms, and explain why it is natural to have a $\chi \epsilon_0$ factor in front of the $F'_{1,2}$ couplings, similar to the appearance of $\chi \epsilon_0$ in front of the flexoelectric $F_{1,2}$ couplings.

A simple way to model the essential physics of a nearly ferroelectric system is to consider it as having one negative and one positive effective ion per unit cell. We then define the "bare" variable $\widetilde{\mathbf{u}}$ to be the average displacement of the two effective ions, and $\widetilde{\mathbf{w}}$ the relative displacement between them. The action generically takes the form:

$$S = \int d^3\mathbf{r} \left[ \frac{\rho_+}{2} \left( \dot{\widetilde{\mathbf{u}}}(\mathbf{r}) + \frac{\dot{\widetilde{\mathbf{w}}}(\mathbf{r})}{2} \right)^2 + \frac{\rho_-}{2} \left( \dot{\widetilde{\mathbf{u}}}(\mathbf{r}) - \frac{\dot{\widetilde{\mathbf{w}}}(\mathbf{r})}{2} \right)^2 - \frac{\rho_e^2}{2(\epsilon - \epsilon_\infty)} \widetilde{\mathbf{w}}^2(\mathbf{r}) - \frac{\widetilde{K}_1}{2} (\nabla \widetilde{\mathbf{u}}(\mathbf{r}))^2 - \frac{\widetilde{K}_2}{2} (\nabla \cdot \widetilde{\mathbf{u}}(\mathbf{r}))^2 \right.$$
$$\left. - \frac{\widetilde{K}'_1}{2} (\nabla \widetilde{\mathbf{w}}(\mathbf{r}))^2 - \frac{\widetilde{K}'_2}{2} (\nabla \cdot \widetilde{\mathbf{w}}(\mathbf{r}))^2 + \rho_e \widetilde{F}_1 \widetilde{\mathbf{w}}(\mathbf{r}) \cdot \nabla^2 \widetilde{\mathbf{u}}(\mathbf{r}) + \rho_e \widetilde{F}_2 \widetilde{\mathbf{w}}(\mathbf{r}) \cdot \nabla(\nabla \cdot \widetilde{\mathbf{u}}(\mathbf{r})) + \ldots + \rho_e \frac{\mathbf{B}}{c} \cdot \left( \widetilde{\mathbf{w}}(\mathbf{r}) \times \dot{\widetilde{\mathbf{u}}}(\mathbf{r}) \right) \right]$$
$$- \int d^3\mathbf{r} d^3\mathbf{r}' \frac{\rho_e^2}{4\pi \epsilon_\infty} \frac{(\nabla \cdot \widetilde{\mathbf{w}}(\mathbf{r}))(\nabla' \cdot \widetilde{\mathbf{w}}(\mathbf{r}'))}{|\mathbf{r} - \mathbf{r}'|} . \tag{29}$$

Note that by our definitions of $\widetilde{\mathbf{u}}$ and $\widetilde{\mathbf{w}}$, the $\mathbf{B}$ term and Coulomb term are exact without higher order corrections. Now, even at $\mathbf{k} = 0$, there is the $\dot{\widetilde{\mathbf{u}}} \cdot \dot{\widetilde{\mathbf{w}}}$ mixing if the effective ion masses are different, $\rho_+ \neq \rho_-$. Removal of such mixing amounts to redefine a decoupled basis: $\mathbf{u}(\mathbf{k})$ and $\mathbf{w}(\mathbf{k})$. At $\mathbf{k} = 0$, the definition of $\mathbf{u}(\mathbf{k} = 0)$ is the displacement of the center of mass (although there is no "acoustic phonon mode" at $\mathbf{k} = 0$, it is still useful to define this displacement for later discussions), while $\mathbf{w}(\mathbf{k} = 0) = \widetilde{\mathbf{w}}(\mathbf{k} = 0)$. This change of variable will readily induce a $(\mathbf{B}/c) \cdot (\rho'_e \mathbf{w} \times \dot{\mathbf{w}})$ term where $\rho'_e$ is proportional to $\rho_+ - \rho_-$. Next we consider $\mathbf{k} \neq 0$. The change of variables from $\widetilde{\mathbf{u}}(\mathbf{k}), \widetilde{\mathbf{w}}(\mathbf{k})$ to $\mathbf{u}(\mathbf{k}), \mathbf{w}(\mathbf{k})$, on top of the said change at $\mathbf{k} = 0$, also includes finite $\mathbf{k}$ corrections which must be proportional to the ratio between $k$ and the gap which is in turn proportional to $1/\sqrt{\epsilon - \epsilon_\infty}$. Consequently, the coupling to $\mathbf{B}$ develops a $k^2$ dependence with coefficients proportional to $\epsilon - \epsilon_\infty$, which for large $\epsilon$ is approximately $\chi \epsilon_0$. This explains why the flexoelectric effect of acoustic phonons, as well as the analogous term for optical phonons, are approximately proportional to $\chi$. Moreover, it is easy to see that typically the signs of $F'_{1,2}$ are opposite to the sign of $F_{1,2}$, with $|F'_{1,2}| \lesssim |F_{1,2}|$. Finally, we make a few side remarks on the Coulomb interaction term. Upon the change of variables, all the mixing terms cancel by the definition of $\mathbf{w}(\mathbf{k})$ and $\mathbf{u}(\mathbf{k})$, including those from the Coulomb interaction term. In addition, as an approximation, we neglect the Coulomb interaction term between two acoustic phonons, which is at the order of $k^4$ if we wrote it in terms of $\mathbf{u}(\mathbf{k})$ and is much smaller than the other terms at low energy. Therefore, the Coulomb interaction between optical phonons written in Eq. (22) is the only term that is important at low energy.

Now we compute the contributions from the couplings in Eq.(28) to $\kappa_H$. As they are entirely independent from the acoustic phonons, we can restrict our attention to the $\mathbf{w}$ phonons only. Also, since we notice that the longitudinal phonon gap can be as large as 1000K [13] in these nearly ferroelectric instulators, we are justified in keeping only the softened transverse modes. To be more explcit, we present the intrinsic thermal Hall contributions from the $\rho'_e$ term and the $F'_{1,2}$ terms individually, since their contributions are additive at linear order in $\mathbf{B}$. First, we discuss the contribution from the $\rho'_e$ term. The energies are modified by

$$\delta^{(1)} E'_{1,2}(\mathbf{k}) = \pm \frac{\rho'_e \mathbf{B} \cdot \mathbf{k}}{\rho' c \hbar k}. \tag{30}$$

The splitting between $E'_{1,2}$ determines a basis for the $\alpha = 1, 2$ modes, between which there is Berry curvature

$$\mathbf{\Omega}^{(0)'}{}_{1,2}(\mathbf{k}) = \mp \hbar \frac{\mathbf{k}}{k^3}. \tag{31}$$

There is a contribution to $\mathbf{\Omega}$ that is directly proportional to $\rho'_e$ as well, which can be computed, but is rather complicated. In the limit of a nearly ferroelectric insulator, where we can treat $v'_T k$, $v'_L k$, $\Delta_{\mathrm{TO}} \ll \Delta_{\mathrm{LO}}$ the Berry

curvature correction is given by the simpler expression:

$$\delta^{(1)}\boldsymbol{\Omega}'_{1,2}(\mathbf{k}) = \frac{\rho'_e}{2\rho'ck^4(E'_T)^3}\left[{v'_T}^2 k^4 \mathbf{B} - (2\Delta_{\text{TO}}^2 + 3{v'_T}^2 k^2)(\mathbf{B}\cdot\mathbf{k})\mathbf{k}\right]. \tag{32}$$

The contribution to the intrinsic thermal Hall conductivity from the $\rho'_e$ term is

$$(\kappa'^{in}_H)^{(1)} = -\frac{1}{3\pi^2}\left(\frac{T}{\hbar v'_T}\right)\frac{\rho'_e B}{\rho' c} R_1(\Delta_{\text{TO}}/T)\left(\frac{\Delta_{\text{TO}}}{T}\right)^2, \tag{33}$$

where the functions $R_n(x)$ are defined as

$$R_n(x) \equiv \int_x^\infty d\xi\ \xi^n \frac{1}{\sqrt{\xi^2 - x^2}}\left(\frac{-\partial f_{BE}(\xi)}{\partial \xi}\right). \tag{34}$$

The behavior of these functions at $x = 0$ is $R_n(x=0) = (n-1)!\zeta(n-1)$, with $\zeta(x)$ the Riemann zeta function. This limit corresponds to an intermediate temperature regime when $T$ is larger than $\Delta_{\text{TO}}$ but still much smaller than the upper edge of the transverse optical phonon band. Notably, in this limit, the behavior of $(\kappa'^{in}_H)^{(1)}$ is

$$(\kappa'^{in}_H)^{(1)} \approx \frac{\pi}{2}\left(\frac{\Delta_{\text{TO}}}{\hbar v_T}\right)\frac{\rho'_e B}{\rho' c}, \tag{35}$$

which could lead to a temperature "plateau" of thermal Hall conductivity in the said intermediate temperature regime. On the other hand, in the large $x$ limit (low temperature compared to the optical phonon gap), the Bose-Einstein distribution in these functions can be approximated by a Maxwell-Boltzman distribution $e^{-\xi}$ and the optical phonon contribution is exponentially suppressed by its gap.

Next we move to the contributions to the intrinsic thermal Hall conductivity from the $F'_{1,2}$ terms. The corrections from these terms to the energies and to the Berry curvatures are given by

$$\delta^{(2)}E'_{1,2}(\mathbf{k}) = \mp\frac{\epsilon F'_1 k \mathbf{B}\cdot\mathbf{k}}{\rho' c \hbar} \tag{36}$$

(the Berry curvature due to this lifting of degeneracy has already been account for by ${\boldsymbol{\Omega}^{(0)}}'_{1,2}(\mathbf{k})$ in the above) and in the limit of $v'_T k,\ v'_L k,\ \Delta_{\text{TO}} \ll \Delta_{\text{LO}}$, we have

$$\delta^{(2)}\boldsymbol{\Omega}'_{1,2}(\mathbf{k}) = \frac{\chi\epsilon_0(2F'_1 + F'_2)}{4\rho'c(E'_T)^3}\left[({v'_T}^2 k^2 + 2\Delta_{\text{TO}}^2)\mathbf{B} + {v'_T}^2(\mathbf{B}\cdot\mathbf{k})\mathbf{k}\right]. \tag{37}$$

The intrinsic thermal Hall conductivity due to the $F'_{1,2}$ terms is

$$(\kappa'^{in}_H)^{(2)} = \left(\frac{T}{\hbar v'_T}\right)^3 \frac{\chi\epsilon_0 B}{\rho' c}\left[-\frac{1}{6\pi^2}G_3(\Delta_{\text{TO}}/T)\,F'_2 + \frac{1}{3\pi^2}G_1(\Delta_{\text{TO}}/T)\left(\frac{\Delta_{\text{TO}}}{T}\right)^2\left(\frac{2F'_1 + F'_2}{2}\right)\right],$$

where the functions $G_n(x)$ are defined as

$$G_n(x) \equiv \int_x^\infty d\xi\ \xi^n \sqrt{\xi^2 - x^2}\left(\frac{-\partial f_{BE}(\xi)}{\partial \xi}\right) = R_{n+2}(x) - x^2 R_n(x). \tag{38}$$

The behavior of these functions at $x = 0$ is $G_n(x=0) = (n+1)!\zeta(n+1)$. Note that in this intermediate temperature regime (when $T$ is larger than $\Delta_{\text{TO}}$ but still much smaller then the upper edge of the transverse optical phonon band), the contributions from $F'_{1,2}$ has a $T^3$ scaling similar to that of the acoustic phonon contribution due to $F_{1,2}$ in Eq.(20). Again, in the large $x$ limit, we can approximate the Bose-Einstein distribution by a Maxwell-Boltzmann distribution. The total thermal Hall conductivity should have additive contributions from both acoustic phonons and optical phonons, including Eq.(20), Eq.(33) and Eq.(38).

In STO, there is indeed a single softened optical phonon branch with gap $\Delta_{\text{TO}} \approx 24\text{K}$ [14]. While this is above the temperature range $T \lesssim 10\text{K}$ on which we focused in the Main Text, the relevant finite temperature factors $G_n(\Delta_{\text{TO}}/T)$ and $R_n(\Delta_{\text{TO}}/T)$ have not decayed to small values; for instance, $G_3(\Delta_{\text{TO}}/T)$ at $T = 10K$ is about $0.7G(0)$. Therefore, we have explicitly computed the effects of the optical phonons here. Assuming $\rho'_e \lesssim \rho_e = 6e/a^3$,

$\rho' \lesssim \rho$, $K'_{1,2} \sim K_{1,2}$ and $F'_{1,2} \sim -(\rho'/\rho)F_{1,2}$, we can estimate that, at $T = 10$K, the $\rho'_e$ contribution to $\kappa_H^{in}$ is about $0.006\text{sgn}(F)\,(\rho'_e/\rho_e)(\rho'/\rho)^{-1/2}$ of the acoustic phonon contribution, and the $F'$ contribution is about $-0.03(\rho'/\rho)^{3/2}$ of the acoustic phonon contribution.

---

[1] T. Qin, J. Zhou, and J. Shi, Physical Review B **86**, 104305 (2012).
[2] X. Li, B. Fauqué, Z. Zhu, and K. Behnia, Physical Review Letters **124**, 105901 (2020).
[3] N. Cooper, B. Halperin, and I. Ruzin, Physical Review B **55**, 2344 (1997).
[4] T. Qin, Q. Niu, and J. Shi, Physical review letters **107**, 236601 (2011).
[5] J. Scott and H. Ledbetter, Zeitschrift für Physik B Condensed Matter **104**, 635 (1997).
[6] J. Dec, W. Kleemann, and B. Westwanski, Journal of Physics: Condensed Matter **11**, L379 (1999).
[7] P. Zubko, G. Catalan, A. Buckley, P. Welche, and J. Scott, Physical Review Letters **99**, 167601 (2007).
[8] P. Zubko, G. Catalan, A. Buckley, P. Welche, and J. Scott, Physical Review Letters **100**, 199906(E) (2008).
[9] P. Zubko, G. Catalan, and A. K. Tagantsev, Annual Review of Materials Research **43** (2013).
[10] S. M. Kogan, Soviet Physics-Solid State **5**, 2069 (1964).
[11] J. Hong, G. Catalan, J. Scott, and E. Artacho, Journal of Physics: Condensed Matter **22**, 112201 (2010).
[12] R. H. Lyddane, R. G. Sachs, and E. Teller, Phys. Rev. **59**, 673 (1941).
[13] J. Ruhman and P. A. Lee, Phys. Rev. B **94**, 224515 (2016).
[14] S. Rowley, L. Spalek, R. Smith, M. Dean, M. Itoh, J. Scott, G. Lonzarich, and S. Saxena, Nature Physics **10**, 367 (2014).



\end{document}